\documentclass{frontiersFPHY} 

\usepackage[english]{babel}
\usepackage{hyperref}

\def\keyFont{\fontsize{8}{11}\helveticabold }
\def\firstAuthorLast{Fraix-Burnet {et~al.}} 
\def\Authors{Didier Fraix-Burnet\,$^{1,2,*}$, Marc Thuillard\,$^{3}$, Asis Kumar Chattopadhyay\,$^{4}$}
\def\Address{$^{1}$Univ. Grenoble Alpes, IPAG, F-38000 Grenoble, France  \\
$^{2}$CNRS, IPAG, F-38000 Grenoble, France \\
$^{3}$La Colline, 2072 St-Blaise, Switzerland \\
$^{4}$Department of Statistics, University of Calcutta, India}
\def\corrAuthor{Didier Fraix-Burnet}
\def\corrAddress{Institut de Plan\'etologie et d'Astrophysique de Grenoble (UMR 5274),
BP 53,
F-38041 Grenoble C\'edex 9,
France}
\def\corrEmail{didier.fraix-burnet@obs.ujf-grenoble.fr}

\newcommand{\mnras}[1]{Monthly Notices of the Royal Astronomical Society}
\newcommand{\apj}[1]{The Astrophysical Journal}
\newcommand{\apjl}[1]{The Astrophysical Journal Letters}
\newcommand{\apjs}[1]{The Astrophysical Journal Supplement Series}
\newcommand{\aj}[1]{The Astronmical Journal}
\newcommand{\aap}[1]{Astronomy \& Astrophysics}
\newcommand{\aaps}[1]{Astronomy \& Astrophysics Supplement Series}
\newcommand{\araa}[1]{Annual Review Astrononmy \& Astrophysics}

\begin{document}
\onecolumn
\firstpage{1}

\title[Multivariate classification of galaxies]{Multivariate Approaches to Classification in Extragalactic Astronomy}
\author[\firstAuthorLast ]{\Authors}
\address{}
\correspondance{}
\extraAuth{}
\topic{}

\maketitle


\begin{center}
2015, \href{http://www.frontiersin.org/milky_way_and_galaxies/10.3389/fspas.2015.00003/abstract}{Frontiers in Astronomy and Space Sciences} 2, 3 \end{center}

\begin{abstract} 

\section{}

Clustering objects into synthetic groups is a natural activity of any science. Astrophysics is not an exception and is now facing a deluge of data. For galaxies, the one-century old Hubble classification and the Hubble tuning fork are still largely in use, together with numerous mono- or bivariate classifications most often made by eye. However, a classification must be driven by the data, and sophisticated multivariate statistical tools are used more and more often. In this paper we review these different approaches in order to situate them in the general context of unsupervised and supervised learning. We insist on the astrophysical outcomes of these studies to show that multivariate analyses provide an obvious path toward a renewal of our classification of galaxies and are invaluable tools to investigate the physics and evolution of galaxies.

\tiny
 \keyFont{ \section{Keywords:} Clustering --  Classification -- Galaxies -- Multivariate Analysis -- Phylogenetic Methods} 
\end{abstract}


\section{Introduction}

Astrophysics has always adopted specific strategies to classify a relatively modest amount of diversity and has much counted on the physics to define the discriminant parameters. This discipline is now facing the need for sophisticated statistical tools to tackle the astronomical number of observed and catalogued objects and the increasing number of observed properties that describe them.

The debate about the usefulness of the morphological classification of galaxies is a rather old one and is still alive. Sandage \citep{Sandage2005}, a proponent of a (morphological) classification driven by the data, noticed that the Hubble classification and the Hubble tuning fork have not yet been replaced by anything else despite the efforts of the proponents of a classification driven by the physics \citep[e.g.][]{Conselice2006}. It also has been recognised to have many flaws: it is a qualitative, subjective and visual approach, difficult to use for distant galaxies, it is based solely on the visible morphological parameter while galaxies are complex and evolving systems and while we have at our disposal morphologies from X-rays to radio wavelengths, spectra, chemical compositions, stellar populations, central black hole masses, kinematics of stars and gas...

However this debate may not address the right question since from a classification point of view, a classification must be driven by the data, and thus be multivariate \cite[e.g.][]{Fayyad1996}. Consequently, adapted tools must be used which are not well known to astronomers in general. Nevertheless, numerous studies have been published during the last thirty years or so, especially since the beginning of the XXIst century. In this paper we would like to present these different approaches in the general context of unsupervised (clustering) and supervised (classification) learning.

Clustering approaches gather objects according to their similarities either through the choice of a distance metric or using some adequate criteria for deciding to which cluster some object belongs. There is a huge class of techniques that partition the data into a pre-defined number of clusters. A well-known algorithm is the k-means \citep{kmeans1967,kmeans2010}.

Another family of clustering techniques uses a hierarchical representation of the pairwise distances between objects in terms of a number of parameters (variables), through a bottom-up algorithm that constructs a tree by relating the closest objects together before relating these first clustering to closest clusters or objects, and so forth until the whole sample is exhausted. The final number of groups is then chosen by cutting the tree at a fixed distance level.
The branches of the tree, called a dendrogram, may or may not represent relationships between the objects.

Originally, phylogenetic methods are designed to build a graph representing the evolutionary relationships between species \citep[see reviews in ][]{Felsenstein2003,Makarenkov2006}. Each node of the graph indicates a transmission with modification mechanism that creates two or more species inheriting from a common ancestor.
More generally, a phylogenetic approach can be viewed as an unsupervised clustering approach in which relationships are provided. As a  consequence, phylogenetic techniques are particularly versatile and powerful methods for building classification trees. They can be understood in the framework of the graph theory \citep{semple2003}.

There are two kinds of phylogenetic methods, based either on the pairwise distances (or dissimilarities) computed from the parameters describing the objects, or on these parameters themselves.

The distance-based methods build the tree entirely from the distances, putting forward the global similarities between the objects. The friends-of-friends algorithm is relatively famous in astrophysics \citep[e.g.][and references therein]{More2011}. Also known as the single linkage or Nearest Neighbor algorithm, it is mathematically related to the Minimum Spanning Tree technique which looks for the simplest graph connecting the objects under study \citep{Gower1969,FeigelsonBabu2012}. A more sophisticated approach used in phylogenetic studies is the Neighbor-Joining Tree technique \citep{NJ1987,NJ2006}.

In the parameter-based methods, the parameters are called characters which in astrocladistics correspond to the parameters associated to the physical measurements of some properties of the objects. The parameter-based methods evaluate all possible trees that can be constructed with the objects, and select the tree(s) corresponding to an optimization criterion. The process is thus based on the distribution of the parameter values.

Parameter-based methods can describe a larger variety of evolutionary scenarios and are thus more general that the distance-based methods. But this is at the cost of a larger computation time which quickly becomes prohibitive. Mathematically, formal connections between parameter- and distance-based methods are developed in the case of continuous parameters \citep[e.g.][]{TF09, TF15}, explaining why both kinds of methods are successfully used in phylogenetic studies.

Among the parameter-based techniques, cladistics is the most famous one. Invented in the 1950's by William Hennig~\citep{hennig1965}, its principle looks simple: two (or more) objects are related if they share a common history, that is they possess properties inherited from a common ancestor. In practice, a cladistic analysis asks for the objects under study to be described by evolutionary characters (parameters or descriptors) for which at least two states are defined: one is said to be ancestral, the other one is said to be derived. The derived state corresponds to an innovation in the evolution and is assumed to have been acquired by an unidentified ancestor. This is the transmission phase of inheritance making descendants look similar to their parents. The accidents in this process are called modifications and generate diversity. This transmission with modification process was invoked by Darwin to explain the observed hierarchical organisation of the biological diversity. Several approaches have been developed to search for the best tree representation using Maximum Likelihood, some Bayesian approaches or Maximum Parsimony. In Maximum Parsimony, one searches for the tree representation of the data with the smallest number of evolutionary steps to explain the data. But in essence, any entity, be it biological or not, evolving with a transmission with modification process can be a priori studied by Maximum Parsimony, provided evolutionary states can be defined for the characters.

A more general representation of relationships are given by networks even though their interpretation is quite complex, but they can be approximated by several trees.

In this review, we do not intend to present all possible techniques in both supervised and unsupervised learning. Rather, we focus on the astrophysical published studies made with the objective of discovering structures in a data set, in other words a new clustering and possibly a new classification of galaxies, beyond the traditional Hubble morphological scheme. We refer the reader to the complete review by \citet{Ball2010} on data mining tools used in astrophysics for further information and references in particular on the separation of sources or the classification of galaxies into morphological types. Our paper is mainly devoted to unsupervised classification (clustering) and presents the phylogenetic methods which are not included in \citet{Ball2010}. In addition we insist on the astrophysical outcome and the new insights that such studies have brought to our knowledge on galaxy physics and diversity.

Part of this paper is inspired from \citet{De2013} which compares the applicability of some of the clustering techniques on the basis of Gaussian and non Gaussian astronomical data sets. Here we do not make such a comparison.

The paper is organized as follows. The first section presents a frequent prerequisite to data mining, the dimension reduction (Sect.~\ref{reddim}). This approach has been heavily used in the extragalactic literature to identify groups in the reduced component space, the motivation being mainly for automatic classification in large data sets. The second section describes the important difference between this motivation, called (supervised) classification, and the clustering (unsupervised classification) which is the main topic of this paper (Sect.~\ref{supervisedunsupervised}). We also discuss shortly the concept of similarity between objects.

Partitioning methods divide the sample into distinct groups. This can be made with hard or soft bounds depending on whether the membership is a probability or not \citep[see e.g. ][]{Andrae2010}. The $k$-Nearest Neighbor (Sect.~\ref{NN}), Support Vector Machine (Sect.~\ref{SVM}) and k-means (Sect.~\ref{kmeans}) methods are of the first kind.
The fuzzy clustering approach (Sect.~\ref{fuzzy}) belongs to the soft partitioning techniques and often extends the applicability of the previous methods. The Information Bottleneck approach is able to provide both kinds of classification (Sect.~\ref{information}).

These partitioning methods require the number of classes as an input. Some other techniques try to fit some distributions to the data set, the optimization process providing the number of groups best fitting the data. These techniques are based on mixture model (Sect.~\ref{mixturemodels}) and wavelet (Sect.~\ref{wavelet}) methods.

A different category of clustering approaches establishes relationships between the objects and derive the groups from the generated graph. The first such category are the hierarchical methods (Sect.~\ref{hierarchical}) which build a tree based on the pairwise distances. Different cuts on the tree result in different numbers of classes. These cuts can be chosen on the basis of objective arguments but also may vary according to the goal of the analysis since the tree provides a synthetic view of the structures within the data set, instead of just the group memberships. Another kind of graphs are the networks produce by the Minimum Spanning Tree method (Sect.~\ref{MST}). The last kind of relationships are evolutionary relationships. This is the domain of the phylogenetic techniques, a very wide subject of bioinformatics. We here present only the Maximum Parsimony (cladistics), Neighbor-Joining Tree Estimation and Outer Planar Networks that have been applied in the context of galaxies (Sect.~\ref{meth:phylogenetic}).

\section{Dimension reduction approaches}
\label{reddim}

\subsection{Methods}

When the data set is large (both in terms of number of variables and number
of observations) one may first apply some appropriate dimension reduction
technique and then perform clustering on the reduced data set.

One must keep in mind that the discriminant usefulness of distances is lost in high dimension parameter spaces since distances tend to become similar (one of the aspects of the ``curse of dimensionality'').

\textbf{Principal component analysis (PCA)}

In this technique, given a data set of observations on correlated variables, an orthogonal transformation is performed
to convert it into a set of uncorrelated variables called the principal components. The number of principal components is less than or equal to the
number of original variables. This transformation is defined in such a way
that the first principal component has the largest possible variance.
One rule of thumb is to consider those components whose eigen values are greater than one in the reduced space.
Principal components are guaranteed to be independent only if the variables are jointly normally distributed.

\textbf{Independent component analysis (ICA)}

Principal component analysis, Factor Analysis, Projection Pursuit are some
popular methods based on linear transformation. But ICA is different because it looks for the components in the representation
that are both statistically independent and non Gaussian. ICA separates
statistically independent components, which are the original source data,
from an observed set of data mixtures. All information in the multivariate
data sets are not equally important. There is often a need for extraction of
the most useful information. ICA extracts and reveals useful hidden factors
from the whole data sets. ICA defines a generative model for the observed
multivariate data, which is typically given as a large database of samples. Contrarily to PCA, the components are not imposed to be orthogonal.

Independent Component Analysis \citep{Comon1994}, model assumes the form
\begin{equation}
 X = AS
 \label{eq:ICA1}
\end{equation}

where $X$ is a data matrix, $A$ is the non-singular mixing matrix, $S$ is matrix
of independent components. $A^{-\vert}$ is the unmixing matrix. The main goal of
ICA is to estimate the unmixing matrix $A^{-\vert}$ . It is assumed that the data
variables are linear or non-linear mixtures of some latent variables and the
mixing system of equation~\ref{eq:ICA1} can be written as
\begin{equation}
X_i = a_{i1} S_1 + a_{i2} S_2 + ...... + a_{in} S_n , i = 1, 2, ..., n
 \label{ICA2}
\end{equation}

The $S_i$ are mutually independent and $a_{ij}$ are the entries of the non-singular
matrix $A$. Here $n$ is the number of parameters (variables). For performing ICA, the data set has to be whitened in the sense
that correlations in the data have to be removed.

There is no rigorous method to determine the optimum number of ICs. For instance, the
number of independent components can be taken to be equal to the number of principal components with eigen values greater than 1 \citep{Albazzaz2004}.
As most of the data sets in Astrophysics are likely to be non Gaussian, ICA
can be successfully used in many situations \citep{Chattopadhyay2013a,Chattopadhyay2013b}.

\subsection{Applications}

PCA technique was applied in a few papers in the 1970s and 1980s with the goal of finding the main parameters explaining the variance among galaxy samples.
For instance, \citet{Watanabe1985} used four parameters (diameter, magnitude, mean surface brightness and mean concentration index) and found that two principal components explains 97\% of the total variance in their sample of all morphological types, in agreement with other studies. While \citet{Watanabe1985} do not find differences in the two-dimension PC plane between elliptical and disk galaxies, \citet{Whitmore1984} more explicitly looks for an objective classification of galaxies: `` The fact that there are so many different classification systems for galaxies...demonstrates that we are still searching for the fundamental properties.''. Using more parameters (up to 15) they agreed with the other studies on two components explaining most of the variance, and tentatively identify them as scale and form. They do not devise a new classification scheme, but rather identify different correlations depending on the position of the galaxies on the 2D diagram.

\citet{Chattopadhyay2006} also found two components in a PCA analysis of samples of spiral galaxies with extended rotation curves. They constructed new ``fundamental planes'' with these components, pinpointing the most important physical factors. They also performed a multiple stepwise regression analysis of the variation of the overall shape of the rotation curves and find that it is mainly determined by the central surface brightness, while the shape purely in the outer part
of the galaxy (beyond the optical radius) is mainly determined by the size of the galactic disk. Such a regression is interesting to predict still unobserved values for some parameters, and is improved by the reduction of the dispersion in the principal component space.

\citet{Peth2015} used PCA as a simple way to reduce the dimensionality, break internal degeneracies and find the natural distributions of data in the parameter space characterizing the structures and shapes of galaxies that they study. These principal components are then used to classify the shape of galaxies through a hierarchical clustering technique (see Sect.~\ref{hierarchical}).

Several studies \citep[e.g.][]{Connolly1995} used PCA both as a dimension reduction and as a tool for classification of spectra of galaxies. Spectra are characterized by a high number of attributes (the wavelengths) that are not independent since a spectrum is made of a continuum spectrum from stars plus absorption and emission lines from the gas. PCA has in principle the power to identify the minimum number of spectra to combine in order to obtain the observed diversity. \citep{Connolly1995} used a variant of the PCA technique, the Karhunen-Loève transform, which allows for weighting differently some parts of the spectra. They not only find that two eigenspectra are necessary to account for most of the variance of the spectra of galaxies, but the distribution of classes in the two-parameter space is one-dimensional. They proposed a scheme of ten classes, some corresponding to the broad morphological types Sa, Sb, S0 and E, while the six others are starburst objects. Their work was intended to be used by spectral surveys to classify automatically the observations.

In a similar scope of general classification of galaxies, one must mention the attempt by \citet{Scarlata2007} to build a morphological automated classification of galaxies, the ZEST catalog, using PCA \citep[see][]{Coppa2011} but the parameters used are criticized by \citet{Andrae2010}. This illustrates the importance of the selection of the parameters for a multivariate clustering or classification analysis which at some point may appear arbitrary and subjective. A special care should be brought to this initial step through the analysis of the data set itself with dedicated data mining tools.

Another instance is the classification established by \citet{Conselice2006} using a PCA analysis together with a Spearman Rank correlation test to better understand the parameters of the data set. His approach is to use the PCA on some set of parameters and then understand the physics of the principal components. So the PCA shed light on the underlying physics from which a classification scheme can be built. He finds three dimensions for this scheme, with the mass (scale), the star formation (spectral type) and the interactions/mergers (degree of dynamical disturbances). This should remind that PCA is not a clustering technique per se, it provides a new representation of the data from which a clustering may be performed. Indeed the work by \citet{Conselice2006} proposes new relationships between the morphological classes. His scheme appears as a more physical replacement of the 2D Hubble tuning fork diagram.

The Principal Component Analysis assumes a linear combination of the parameters, a rather strong assumption. \citet{TaghizadehPopp2012} have used a non-linear PCA, the Principal Curve analysis, ``which can be seen as a nonparametric extension of linear PCA. The principal curve is the curve following the location of the local mean in the multi-dimensional cloud of data points.'' They obtain a drastic dimension reduction with a one-dimension parameter space (the Principal Curve) which they divide arbitrarily into 20 groups of equal densities. They compute a distance (the arc length) that ranks the galaxies so providing “a natural and objective way of
ordering, partitioning and classifying the rich zoo of galaxies in the nearby universe”. \citet{TaghizadehPopp2012} do
not include luminosity nor mass in the process in order not to bias the study of the luminosity function
as a function of the arc length. This is debatable but they are right in saying that it would induce a bias
since these parameters will define a strong axis of variance in the PCA. Nevertheless would it be possible to
classify galaxies without their mass? Could massive galaxies have the same history as less massive
ones? This shows that the choice of the parameters is never so obvious, and generally related to the choice of the technique used as well. The interesting point is that they recover known trends in the physics of galaxies, but more importantly they can identify new kinds of galaxies pointing out particular physical processes and histories of galaxies. These discoveries can only be made by multivariate analyses.

\citet{Folkes1996} applied PCA on spectra of low signal-to-noise ratio mainly as a dimensionality reduction technique. The few principal components are then used to train a neural network in order to classify galaxies into the five broad morphological types. Even though this approach is efficient for big data sets, it appears limited to normal galaxies since they find that a new classification scheme must be used where unusual features are present in the spectra.

The ICA analysis is still less common than PCA for the study of galaxies. At least two studies have been published, an ensemble learning for ICA \citep{Lu2006} and a mean field independent component analysis \citep{Allen2013}. In the first case, 1326 synthetic spectra have been used coming from Single Stellar Population models. They select 74 "sufficiently" different spectra from these (using an objective criterion) since the ensemble learning part converges very slowly. The ICA analysis yields six most significant components, and the 1326 spectra are fitted on these components. Each component represent a basic element behind the spectra of galaxies, and they find that each of them can be associated closely to one or a few stellar types plus some peculiar line properties. These six components are then used on real galaxy spectra to derive the stellar contents like starlight reddening, stellar velocity dispersion, stellar content, and star formation history. Even though PCA is much faster, it does not provide this important information because of the orthogonality constraint that does not allow the components to  be non-negative.

\citet{Allen2013} used the mean field ICA which is a probabilistic ICA using a prior to constrain the components. They find that ten components (divided into five continuum and five emission components) are required to produce accurate reconstructions of essentially all narrow emission-line galaxies to a very high degree of accuracy. Using these ten components on a large sample of Sloan Digital Sky Survey (SDSS) galaxies, they identify the regions of parameter space that correspond to pure star formation and pure active galactic nucleus (AGN) emission-line spectra, and produce high S/N reconstructions of these spectra.

In a similar fashion, \citet{Hurley2014} applied the Non-negative Matrix Factorization technique which has been developed for blind source separation problems. Unlike PCA, this technique imposes the condition that weights and spectral components are non-negative that is also possible in the ensemble learning approach for ICA described above \citep{Lu2006}.
This more closely resembles the physical process of emission in the mid-infrared region studied in this work, resulting in physically intuitive components. They find seven such components, including two for active galactic nucleus emission, one for star formation, and one for the rising continuums at longer wavelengths. They show that the seven components can be used to separate out different types of objects (see Sect.~\ref{mixturemodels}) and to separate out the emission from AGN and star formation regions and define a new star formation/AGN diagnostic which is consistent with all mid-infrared diagnostics already in use but has the advantage that it can be applied to MIR spectra with low signal-to-noise ratio or with limited spectral range.

\section{Supervised and unsupervised learning}
\label{supervisedunsupervised}

\subsection{Distances/dissimilarities}

A lot of learning techniques require a dissimilarity measure. Among them, the distances obey the well-known triangular properties and define a metric. In hierarchical clustering, the distances mainly come from a very general distance known as the
Minkowski's distance or the $p$th norm, which may be defined as follows.
For two points P $= (x_1, x_2 , ....., x_n)$ and Q $= (y_1 , y_2 , ....., y_n)$ in the $n$ dimensional space, the $p$th norm is given by

\begin{equation}
 L_p = \left(\sum_{i=1}^n \vert x_i - y_i \vert^p\right)^{1/p}
\end{equation}

For $p=1$, it gives the Manhattan distance (L$_1$ norm). For $p = 2$, it reduces to the Euclidean distance (L$_2$ norm). Also for $p = \infty$, the L$_p$ norm results
in Chebyshev distance. In hierarchical clustering, Euclidean and Manhattan
distances are mainly used. But these measures are applicable only to continuous data. For categorical or binary data other distances must be used but will not be addressed in this paper.

It may be noted that the selection of the appropriate distance matrix for clustering problems completely depends on the physical situation.

\subsection{Supervised learning (classification)}

Supervised learning technique may be viewed as a mapping between a set
of input variables and an output variable. This mapping is applied
to predict the outputs for unseen data. The main characteristic of supervised learning is the availability of annotated training data. It supervises
the learning system to instruct on the labels to associate with training examples. These labels are known as class labels in classification problems.
Supervised learning induces models for the training data and these models are then used to classify other unlabeled data. Two most popular supervised
learning techniques are the Nearest Neighbor (Sect.~\ref{NN}) and the Support Vector Machines (Sect.~\ref{SVM}) classifiers.

\subsection{Unsupervised learning (clustering)}

The unsupervised learning or clustering seeks some pattern in the data set by starting from the raw data with or without any distributional
assumption regarding the underlying distribution. The three main categories
of this kind are (i) connectivity based clustering (like hierarchical clustering, see Sect.~\ref{hierarchical}), (ii)
centroid based clustering (like k-means, see Sect.~\ref{kmeans}) and (iii) density based clustering (like DBSCAN or more generally kernel density estimation).

An overview of these approaches can be found in \citet{DAbrusco2012} with many references of applications in astrophysics. Most of the methods that we present in the following are unsupervised clustering. The reason is that the multivariate analyses of galaxies essentially are either supervised approaches based mainly on dimension reduction techniques (mostly PCA, see Sect.~\ref{reddim}) or unsupervised methods to discover new classification schemes of galaxies which are really objective and multivariate.

\section{Nearest Neighbor}
\label{NN}

\begin{figure}[h!]
\begin{center}
\includegraphics[width=8cm]{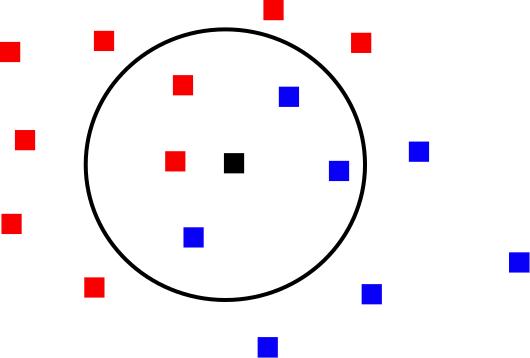}%
\end{center}
 \textbf{\refstepcounter{figure}\label{fig:kNN} Figure \arabic{figure}.}{ In this k-Nearest Neighbor illustration with $k=5$, the central black square more probably belongs to the blue class. }
\end{figure}

\subsection{Method}

The $k$-Nearest Neighbor (NN) algorithm is very intuitive. It starts from a training set for which we have the class labels.  In order to make a prediction about a new observation, one looks at the labels of its $k$ nearest neighbors and uses a majority vote to make the prediction (Fig.~\ref{fig:kNN}).
As the number of neighbors used in making the prediction increases, the decision boundaries become smoother, the bias increases, but the variance decreases.

\subsection{Applications}

\citet{Ball2007} explored the k-Nearest Neighbor technique for determining photometric redshifts in petascale databases using 55~746 quasar spectra from the SDSS. The algorithm is trained on a representative sample of the data. The main result is that the comparison between the photometric and the spectroscopic redshifts shows no region of catastrophic failure where the two derived values differ a lot, contrarily to other methods used to derive photometric redshifts.

\section{Support Vector Machine}
\label{SVM}

\subsection{Method}

Support Vector Machine (SVM) aims to find the hyperplane that best separates two classes of data through an optimization method. Instead of using just a standard
orthogonal basis, SVM uses many functions to describe good separating surfaces. The input data are viewed as sets of vectors, and the data points closest to the classification boundary, determined from a training sample, are the support vectors. SVM fundamentally separates two classes of objects which is probably a limitation in its use for the classification of galaxies.

They use optimization methods to find surfaces that
best separate categories. Their key innovation is to express the separating surfaces in
terms of a vastly expanded set of basis functions. Instead of using just a standard
orthogonal basis, SVMs use many basis elements.

\subsection{Applications}

SVM has been used by \citet{HuertasCompany2008} for the morphological classification of galaxies from the COSMOS survey. The training sample is a limited sample classified visually using a 12-dimensional volume, including 5 morphological parameters, and other characteristics of galaxies such as luminosity and redshift. The objective is to be able to classify automatically the results of big surveys. However, the result seems a little bit disappointing since it can only separate between the two broad classes of early- and late-type galaxies, with an error of about 20\%, even though this is better than other methods generally used.

\section{K-Means}
\label{kmeans}

\subsection{Method}

\begin{figure}[h!]
\begin{center}
\includegraphics[width=8cm]{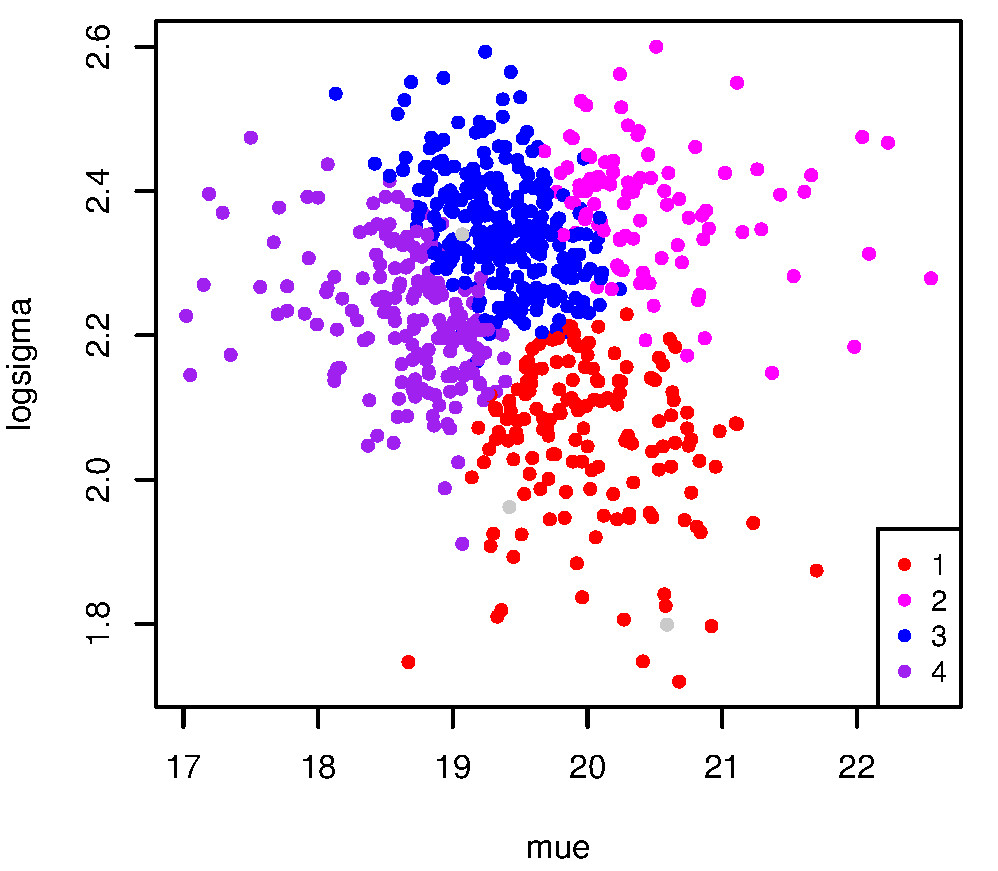}%
\end{center}
 \textbf{\refstepcounter{figure}\label{fig:kmeans} Figure \arabic{figure}.}{ A typical result of a k-means analysis in which the clusters (four here) are clearly distinguishable \citep[from][]{Fraix2010}. }
\end{figure}

The k-means algorithm \citep{kmeans1967,kmeans2010} is a partitioning approach that starts with $k$ centroids, $k$ corresponding to the number of clusters given a priori. It then assigns each data point to the closest centroid and when the clusters are built, the new $k$ centroids are computed and the process iterates until convergence (Fig.~\ref{fig:kmeans}). The result depends very much on the initial centroids. Repeating the analysis with several initial choices is always a good idea, but consistency is not guaranteed if the data do not contain distinguishable and roughly spherical clusters. Some strategies have been devised to guess the best initial choice for the centroids \citep[e.g.][]{Sugar2003,Tajunisha2010} and many indices are available in the package $NbClust$ \citep{NbClust} of R \citep{R}.

A variant called the k-medoids algorithm \citep{kmedoids1987,kmedoids} chooses data points as centers (medoids) and is known to be more robust to noise and outliers.

 \subsection{Applications}

 The k-means algorithm has been used in the context of stars \citep[e.g.][]{Gratton2011,Simpson2012},  galaxies \citep[e.g.][]{Fraix2010,SanchezAlmeida2010,Fraix2012} or Gamma-Ray Bursts \citep{ChattopadhyayGRB2007}.

\citet{SanchezAlmeida2010} performed a k-means analysis of a large number (788~677) of spectra from the SDSS. Each spectra is a collection of about 4~000 wavelengths, making the full data set very computationally demanding for a direct k-means. They thus decided to limit the spectra to a priori informative regions, reducing the number of ``parameters'' to 1637. Their analysis is affected by the dependence of the result on the seed. They say that estimation tools for the number of clusters could not be applied because of the sample size. Using some criteria, they end up choosing randomly one classification having 28 classes. The result looks more like a continuum distribution of spectra, and even if not shown, overlapping between classes is important. This questions the validity of the k-means approach in this case as another k-means analysis of the same sample has shown \citep{De2014}.

Multivariate k-means analyses of smaller sample of galaxies with the aim of discovering new classes of galaxies have been performed as a complement to other clustering methods by \citet{Fraix2010} with the four parameters of the fundamental plane, and by \citet{Fraix2012} with six parameters selected from 23 available. In the latter case, the selection of the parameters is made through different statistical tools, in order to find a parameter subspace in which a robust clustering of the data is present. This leads to the important result that several very different clustering techniques yield compatible clusterings, giving good confidence to the result. The astrophysical implications are numerous since a new classification is established and the average properties and the correlations varies from group to group and often differ from those of the global sample. However, the interpretation benefited from the relationships between the classes established by the phylogenetic method used in these works and discussed in Sect.~\ref{meth:phylogenetic}. Even though the clusters are similar, the absence of these links in the k-means results is clearly missing.

\citet{Chattopadhyay2013} performed a k-means analysis of a large sample of dynamically hot stellar systems from globular clusters to giant ellipticals, in quest of the formation theory of ultra compact dwarf galaxies (UCDs), using three parameters (logarithm of stellar mass, logarithm of effective radius and stellar mass to light ratio). The number of clusters, five, is given by the optimum criterion of \citet{Sugar2003}. The classification of UCDs provides some new clues to the long discussed hypothesis that these objects may be formed either as massive globular clusters or have an origin similar to nuclei of dwarf galaxies.

\section{Fuzzy clustering}
\label{fuzzy}

\subsection{Methods}

In non-fuzzy or hard clustering, data is divided into crisp clusters, where each data point belongs to exactly one cluster. In fuzzy clustering, the data points can belong to more than one cluster, and associated with each of the points are membership grades which indicate the degree to which the data points belong to the different clusters.
Many algorithms exist, many of them being extension of hard clustering algorithms. One example is the fuzzy C-means which is very similar to the k-means (Sect.~\ref{kmeans}) but adding a weight between 0 and 1 to each point characterizing its probability to belong to a given group, and a degree of fuzziness of the groups.

\subsection{Applications}

\citet{Coppa2011} studied the bimodality of galaxies which comes from a double peak distribution in some scatter plots, particularly in color-color diagrams. The origin of this bimodality and the relationship between the two broad classes, ``red'' and ``blue'' or ``late type'' and ``early type'', is still not understood. Evolution is probably involved, but then what is the status of the overlapping regions called ``the green valley''? To know whether this bimodal distribution is an intrinsic property of galaxies and their evolution, multivariate analysis must be used since it appears in several scatter plots. \citet{Coppa2011} use an unsupervised fuzzy partition clustering algorithm applied on the principal components of a PCA analysis. They use eight parameters, two coming from spectra, one from photometry and five describing the morphology. They keep three principal components to perform the clustering analysis which proceeds in two steps: a modified fuzzy k-means algorithm to guess the memberships and the cluster centroids, and a second algorithm (fuzzy modification of maximum likelihood estimation) to achieve optimal fuzzy partition
\citep[see references in][]{Coppa2011}.

They decide to identify three clusters, blue, red and green, somewhat giving up the fuzzy nature of their study. In addition, they name the clusters after previous classifications, even though ``the 'early type cluster is not intended to be made up of pure passive galaxies; rather, it is composed also by bulge-dominated weakly-star forming objects.'' This is a quite confusing practice especially because they discover some new kinds of objects which are invaluable for our understanding of the physics and evolution of galaxies.

Bayesian approaches can also be seen as soft classification as illustrated for instance in the separation between star forming galaxies and Active Galactic Nuclei (AGNs) in \citet{Norman2004} to avoid confusion between different kinds of objects.

\section{Information Bottleneck technique}
\label{information}

\subsection{Method}

The Information Bottleneck Method \citep{Tishby2000} is a simple optimization principle for a model-free extraction of the relevant part of one random variable with respect to another. The algorithm is extremely general and may be applied to different problems in analogous ways. A great advantage of this unsupervised clustering technique is that it avoids the arbitrary choice of the distance and provides a natural quality measure for the resulting classification.

Using the mathematical notations of \citet{Slonim2001} that applied this technique to galaxies, the optimal classification is given by maximizing the functional:
\begin{equation}
 {\mathcal L} [p(c|g)] =  I(C ; \Lambda) - \beta^{-1} I(C; G)
\end{equation}
where $C$ represents the classes, $G$ the galaxy sample and $\Lambda$ the spectral wavelengths. $I(C ; \Lambda)$ and $I(C; G)$ are the mutual information between $C$;$\Lambda$ and $C;G$.
$\beta^{-1}$ is the Lagrange multiplier attached to the complexity constraint.  For $\beta \rightarrow 0$ the classification is non-informative, and for $\beta \rightarrow \infty$ the representation becomes arbitrarily detailed.

\subsection{Applications}

\citet{Slonim2001} explain that by normalizing the total photon counts in each spectrum to unity, we can consider it as a conditional probability, the probability of observing a photon at a specific wavelength from a given galaxy. The ensemble of spectra can thus be seen as a conditional probability distribution function that allows to undertake the information theory-based analysis. For any desired number of classes, galaxies are classified such that the information content about the spectra is maximally preserved.

The number of classes is an issue in most unsupervised clustering techniques, and the information bottleneck shares this difficulty too. \citet{Slonim2001} note that 'the true or correct number of classes may be an ill-defined quantity for real data sets and the number should be determined by the desired resolution, or preserved information``. However one should be careful to use objective arguments based only on statistics, since the physical interpretation should come at the end to tell whether or not the result is interesting.

The main results of this study is the demonstration that an objective and automated technique can yield a classification of spectra which is very physical, in the sense that it recovers results obtained more classically, but is able to discover other classes and correlations between physical parameters. An interesting point in their study is that they applied the same techniques to two samples, one observed and one simulated. The good agreement between the two clusterings shows that the models of galaxy evolution are sensible. This is a good approach to test the models by statistically comparing two populations using multivariate data sets.

\section{Mixture Models}
\label{mixturemodels}

\subsection{Methods}
Most partitioning methods use a distance to define the clusters. In model-based clustering methods, each cluster can be represented by a parametric distribution, the data set being thus considered as a mixture of such model distributions \citep{Qiu2007}. The parameters include the mixing proportions or the prior probabilities of the clusters since the true cluster memberships of the observations are unobserved. The optimization relies on the likelihood of the weighted linear combination of the cluster distributions through the Expectation-Minimization (EM) algorithm. Clustering is done by applying the maximum posterior (Bayes) rule. The process yields a soft classification (probability of membership) and a fit to each cluster distribution.

The mixture model approach also provides expected misclassification probabilities. It requires the number of clusters to be known, which can be for instance estimated with the tools developed for the k-means analysis \citep[][ Sect.~\ref{kmeans}]{Chattopadhyay2009}.

\subsection{Applications}

\citet{Davoodi2006} find four Gaussian distributions best fit the color distribution of 16~698 extragalactic infrared sources. They use this result to propose a classification scheme ($C_a$ to $C_d$) of galaxies that reveal a greater variety of galaxy types than usual spectral energy distribution fitting techniques that strongly depends on the quality of the template model components. Interestingly, \citet{Davoodi2006} use their soft classification to identify outliers (rare galaxies or transient phases) by summing up the four probability density functions for each object.

\citet{Hurley2014} used the seven components they have found with a dimension reduction approach (Sect.~\ref{reddim}) to define a parameter space in which they apply an unsupervised Gaussian Mixture Model clustering algorithm in order to provide a classification tool. This clustering approach is a fuzzy approach since clusters  describe  a probability  density  function indicating how  likely  a  galaxy  could  be  found  in  any  one  of  the  clusters. Eight clusters are found which are consistent with previous classifications. Strangely enough, these clusters are named according to the classical classification through a majority rule. We may ask why use an unsupervised technique if one believes in an existing ''true`` hard classification?

\section{Wavelet Analysis}
\label{wavelet}

\subsection{Method}

The wavelet transform is a well known signal analysis technique widely used in many research areas. Its key property is the ability to provide a multi-resolution approximation of a given input signal through a prototype function $\Psi$:

\begin{equation}
 \label{eq:wavelet}
 W(s,r) = \int f(t) \frac{1}{\sqrt s} \Psi\left(\frac{t - r }{s}\right) dt
\end{equation}
where $s$ characterizes the scale and $r$ the translation factor. The prototype function, also called the mother wavelet, is continuous in both time and frequency and serves as the analysing window.

With this definition, wavelets appear as a parametric-model decomposition of a data set using some basis functions. They could then be used for dimension reduction and/or classification \citep{Thuillard2001}.

Shapelets are  a  scaled  version  of  two-dimensional Gauss-Hermite polynomials and form a set of complete basis functions that are orthonormal on the interval $\left[-\infty,\infty\right]$.
Shapelets are thus suited to decompose images. For galaxies, their use is limited to high signal-to-noise data and rather regular galaxies since they are gaussian-shaped and spherical \citep{Andrae2010}. The composition is an automatic and objective representation of galaxy morphologies.

Other multiresolution methods have been proposed, like for instance the hierarchical Markov models extended for the multispectral astronomical image segmentation \citep{Collet2004}.

\subsection{Applications}

Wavelets can be used to decompose galaxy spectra into several features that can then be used to classify the spectra. In this sense they serve as a dimension reduction technique but contrarily to PCA or ICA the basic elements (features) can be chosen to be physically meaningful, representing the three components of spectra: the continuum, the emission and absorption lines \citep[e.g.][]{Starck1997,Liu2005}.

\citet{Andrae2010} review how an automatic classification of galaxy morphologies could be done using shapelets. Their goal is not to devise a new classification, since it is extremely difficult to parametrise  arbitrary  galaxy  morphologies apart from the question that the morphology is only one property of galaxies. To address the parametrisation problem, they use shapelets and then define the distance as the angle spanned by their (normalised) coefficient vectors of the shapelets:

\begin{equation}
 d(x_i,x_j) = \arccos(x_i\cdot x_j)
\end{equation}

They then use a soft (fuzzy) clustering algorithm with the similarity matrix given by:
\begin{equation}
 W_{mn} \propto \frac{\left(d(x_i,x_j)/d_{max}\right)^{\alpha} }{s}
\end{equation}
with $d_{max}$ being the  maximum  distance  between  any two objects in the given data sample, and $\alpha > 0$  and $s >1$ being  free  parameters  that  tune  the  similarity  measure. This probabilistic clustering technique uses the graph theory in which the similarity elements $W_{mn}$ are the weights of the edges.

They also evaluate the impact of hard clustering methods on the estimation of the parameters characterising the classes depending on the level of overlapping. This is an important point to keep in mind in all hard (non-fuzzy) approaches to clustering, be it by hand or algorithmic. They even suggest that the processes of galaxy evolution and observations tend to invalidate hard clustering approaches.

They do not go into the details of the astrophysical interpretation, but they clearly demonstrate the advantages of such sophisticated approaches for automatic morphological classification of a huge number of galaxies. However, as they rightly say,  ``a lot of work is still needed on the interpretation of the results.''

\section{Hierarchical Classification Methods}
\label{hierarchical}

\begin{figure}[h!]
\begin{center}
\includegraphics[width=8cm]{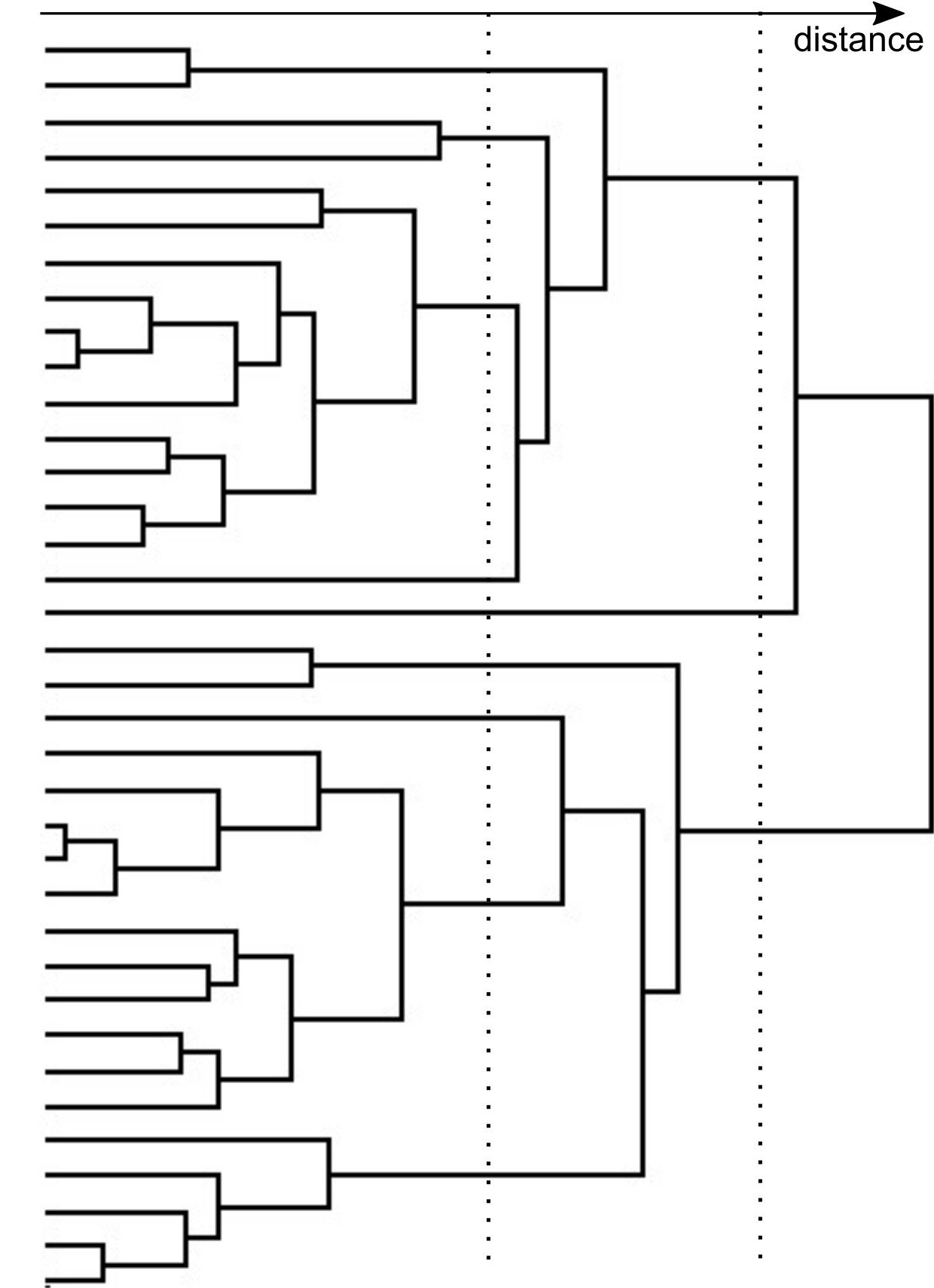}%
\end{center}
 \textbf{\refstepcounter{figure}\label{fig:dendro} Figure \arabic{figure}.}{ An example of a dendrogram. Distance between two horizontal branches is going from left to right. The two dashed lines illustrates two cuts yielding 9 or 3 clusters.}
\end{figure}

\subsection{Methods}

The hierarchical classification method builds a hierarchy of clusters. Two main approaches to form the hierarchy are
agglomerative and divisive. In the agglomerative approach each
observation is considered as a cluster and pairs of clusters are
merged as one moves up the hierarchy (see Fig.~\ref{fig:dendro}). The most similar objects are
grouped first and those initial groups are merged ultimately into
single cluster according to some proximity measure. These proximity
measures are based on either similarities or dissimilarities
(distances). In the divisive analysis approach all observations at first
are grouped in one cluster, and splits are performed recursively as
one moves down the hierarchy. Here an initial single group of
objects is divided into two subgroups such that the objects in one
subgroup are far from the objects in the other. These subgroups
are further divided into dissimilar groups until there are as many subgroups as objects.
%

In order to decide which clusters should be combined or where a
cluster should be split, a distance matrix is required. The
distances used for hierarchical clustering are mainly Euclidean and Manhattan for
continuous type data. In order to find distances between clusters different linkages like
single linkage, complete linkage, average linkage etc are used. Note that the nature
of the final clusters totally depends upon the choices of distances and linkages.

It is interesting to note that if the metric used is the single linkage, then this method is similar to the Minimum Spanning Tree technique (Sect.~\ref{MST}).

\subsection{Applications}

\citet{Peth2015} applied a Ward hierarchical agglomerative clustering to classify galaxies in distinct groups using the first three principal component eigenvectors. In this kind of approach the number of groups is chosen after the analysis. \citet{Peth2015} selected ten groups as a compromise between too many small groups which might appear as too specific, and too large ones that would smear out the true diversity of the objects. They also try to define boundaries to these groups in the PC-space by fitting a convex hull around the points within each groups in order to classify future new observed objects. However, a Nearest-Neighbor or SVM technique could be used in this purpose without the need to compute a convex hull which is a rigid boundary. It is important to recall that a classification is never definitive and would probably evolve with the inclusion of new objects, as it has been for instance the case for the S0 (lenticular) morphological class of galaxies which were not present in the original Hubble classification.

One of the main results of the studies by \citet{Peth2015} is a refined and objective classification of structures and morphologies of the galaxies in their samples. The ten groups are analyzed separately to derive their properties and their probable evolutionary status and history. Their scheme separates quenched compact galaxies from larger, smooth proto-elliptical systems, and star-forming disc-dominated clumpy galaxies from star-forming bulge-dominated asymmetric galaxies. It also reveals a higher fraction of bulge-dominated galaxies than visual classification or one based on the Sersic index.

Decision trees are a practical use of hierarchical clustering.
\citet{SanchezAlmeida2012} propose a decision tree to classify galaxy spectra according to some general features that usually serves as a classification of galaxy properties. They use the decision tree on their previously ASK classes determined with the k-means technique \citep[Sect.~\ref{kmeans}, ][]{SanchezAlmeida2010}. Somehow, in this way, they classify their new classes on another classification.

\citet{Suchkov2005} have applied an oblique decision tree classifier on the homogeneous multicolor imaging data base of the SDSS, the classifier being trained on subsets of objects (stars and galaxies) whose nature is precisely known via spectroscopy. Each node in the decision tree is a criterion on one parameter, defining an hyperplane parallel to one of the axis. In an oblique decision tree, the criterion is based on a (linear) combination of parameters, so the tree is no more parallel to any of the axes in the parameter space. In \citet{Suchkov2005} the classifier is composed of ten oblique decision trees and the final decision is made by votes which yield a class probability distribution for a given object.
The main result of their study is to show the ability of this approach to automatically classify objects from the photometry instead of the spectroscopy which is harder to obtain and analyse, and accurately predict the redshifts of both normal and active galaxies. This can increase considerably the samples required to analyse statistically the evolution and diversity of galaxies, their properties and their correlations.

\section{Minimum Spanning Tree}
\label{MST}

\subsection{Methods}

\begin{figure}[h!]
\begin{center}
\includegraphics[width=8cm]{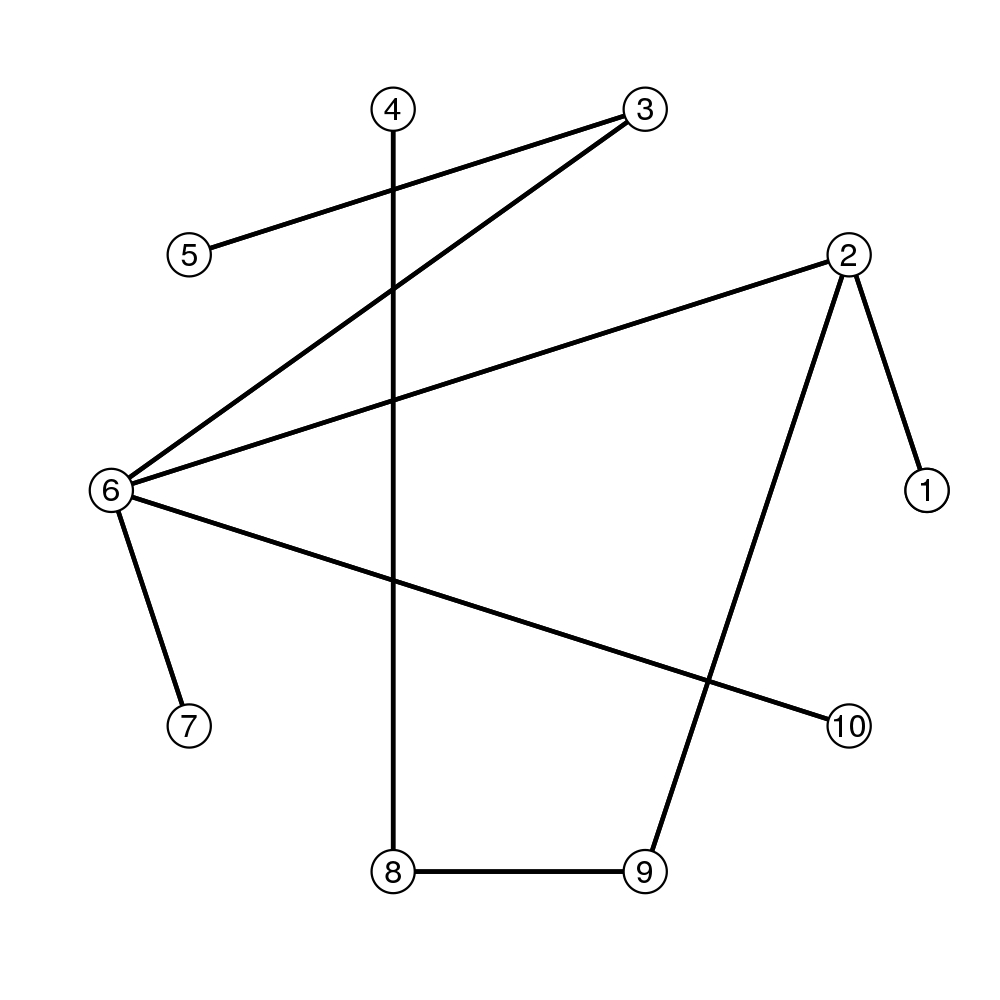}%
\end{center}
 \textbf{\refstepcounter{figure}\label{fig:MST} Figure \arabic{figure}.}{ An illustration of a Minimum Spanning Tree linking ten nodes. }
\end{figure}

The Minimum Spanning Tree (MST) is mathematically related to the single linkage clustering, known to astronomers as the friends-of-friends algorithm or Nearest Neighbor algorithm \citep{Gower1969,FeigelsonBabu2012}. A spanning tree is an acyclic, connected graph $G$ which is a set $(V,E)$ of vertices (nodes) and edges (branches) together with a function $w : E \rightarrow \mathbb{R}$ that assigns a weight $w(e)$ to each edge $e$ in $E$. The minimum spanning tree (Fig.~\ref{fig:MST}), is the spanning tree $T$ minimizing the function :

\begin{equation}
\label{eq:MST}
w(T) = \sum_{e\ \in\ T} w(e)
\end{equation}

If the weights $w(e)$ are distinct, then the solution is unique. A number of algorithms have been developed to solve exactly the Minimum Spanning Tree problem. The first algorithm is attributed to \citet{Boruvka1926}. Other popular algorithms are Prim's, Krukal's and the Reverse-Delete algorithms that all find solutions in polynomial time. The above algorithms also work at higher dimensions in which case the Euclidean L2 or the Manhattan L1 distances are generally used.

Minimum spanning trees have found applications in phylogeny, computer vision, and cytology just to name some domains. It has been used in astrophysics, and maybe very early since a large number of constellations defined by early civilizations are also shown to correlate well with a Minimum Spanning Tree \citep{Dry09}.

\subsection{Applications}

The Minimum Spanning Tree technique has been heavily used to determine the galaxy clusters in order to map the spatial distribution of the baryonic matter, a visible signature of the gravitational structure of the Universe shaped by the Dark Matter \citep{Barrow1985,Bhavsar1996}. This is not an application to clustering in the sense of classification, but this is a spatial clustering. Indeed the MST approach has been strongly adapted to the particular constraint of cosmological observations: the exact position along the line of sight is only approximately given by the redshift. We do not discuss any further this application which is not in the main scope of this paper.

We know of only one use of MST for galaxy classification. \citet{Ascasibar2011} applied this technique to understand their ASK classification of SDSS spectra obtained with the k-means method \citep[Sect.~\ref{kmeans},][]{SanchezAlmeida2010}. They find that the majority of the spectral classes are distributed along a well-defined branch going from the earliest to the latest types, with optically bright active galaxies forming an independent branch that intersects the main sequence exactly at the transition between early and late types. This description is already an interpretation of the 23 ASK classes that present a regular distribution of their spectra as already mentioned in Sect.~\ref{kmeans}, so that the very linear structure of the MST tree is not surprising. However, the approach is interesting because this is a rather simple and objective method to obtain relationships between classes.

\section{Phylogenetic methods}
\label{meth:phylogenetic}

Basically, all galaxies share a common origin which is the gathering of baryonic matter as a self gravitating object. This baryonic matter was very primitive and has subsequently being enriched and diversified by several generations of stars and many transforming processes like galaxy interactions and mergers. There are thus obvious evolutionary relationships between different kinds of galaxies as immediately understood by Hubble when he discovered galaxies and established his famous tuning fork diagram. Taking into account the galaxy diversity of morphologies known at that time, he built a phylogenetic tree in which the relationships are due to the evolution of the stellar orbits which, he thought, should flatten with time because of the dynamical friction. Even though we now know that this process cannot be accomplished in a time shorter than the known age of our Universe, this tuning fork diagram is still used to represent galaxy diversity.

Somewhat strangely enough, phylogenetic analyses of galaxy diversity has not been attempted again for a century. This is probably because the data did not allow much progress into this direction. But we now have huge multivariate databases and it seems timely to reconsider this question. We here present only a few techniques, those that have been already used on astrophysical data sets.

\subsection{Methods}

Before describing some of the most important methods, let us point out that the development of phylogenetic methods has been hindered till the 2000’s by very heated discussions on the philosophical merits of the different approaches. It is only in recent years that most of the barriers between the different schools of thoughts could be overcome by a new generation of researchers. Recently a new picture of phylogenetic methods is emerging. It becomes nowadays increasingly clear that all the different approaches can be discussed within a common framework including distance- and character-based approaches, and that this theoretical framework applies both to phylogenetic trees and networks. 

There are two main categories of methods: the distance-based and the character-based. The ``characters'' are traits, descriptors, observables, parameters or properties, which can be assigned at least two states characterizing the evolutionary stage of the objects for that character. For continuous parameters, these states can be obtained through discretization.

\textbf{Distance-based Approaches: Neighbor Joining Tree Estimation}
\label{meth:NJ}

For distance-based approaches, Neighbor-Joining is the most popular approach to construct a phylogenetic tree.
The Neighbor Joining Tree Estimation \citep[NJ,][]{NJ1987,NJ2006} is based on a distance (or dissimilarity) matrix. This method is a bottom-up hierarchical clustering methods. It starts from a star tree (unresolved tree). A ``corrected'' distance $Q(i,j)$ between objects $i$ and $j$ from the data set of $n$ objects, is computed from the distances $d(i,j)$:

\begin{equation}
\label{eq:NJ}
 Q(i,j) =  (n-2)d(i,j) - \sum_{k=1}^{n}d(i,k)  - \sum_{k=1}^{n}d(j,k)
\end{equation}

The branches of the two objects with the lowest $Q(i,j)$ are linked together by a new node $u$ on the tree. This node replaces the pair $(i,j)$ in the subsequent iterations through the distance to any other object $k$:

\begin{equation}
\label{eq:NJnode}
   d(u,k) = \frac{1}{2}\left[d(i,k)-d(i,u)\right] +  \frac{1}{2}\left[d(j,k)-d(j,u)\right]
\end{equation}

Neighbor-Joining minimizes a tree length, according to a criteria that can be viewed as a Balanced Minimum Evolution \citep{NJ2006}. For a tree metrics, Neighbor-Joining furnishes a simple algorithm to reconstruct a tree from the distance matrix. There is a large literature on how to best approximate a metrics by a tree metrics \citep[see for instance][]{Fakcharoenphol2003}.
Neighbor-Joining is justified if the difference between the original distance matrix and the distance matrix describing the X-tree obtained with Neighbor-Joining is not too large.

\textbf{Character-based Approaches: Cladistics, Maximum Parsimony, Maximum Likelihood...}
\label{meth:chalclad}

\begin{figure}[h!]
\begin{center}
\includegraphics[width=8cm]{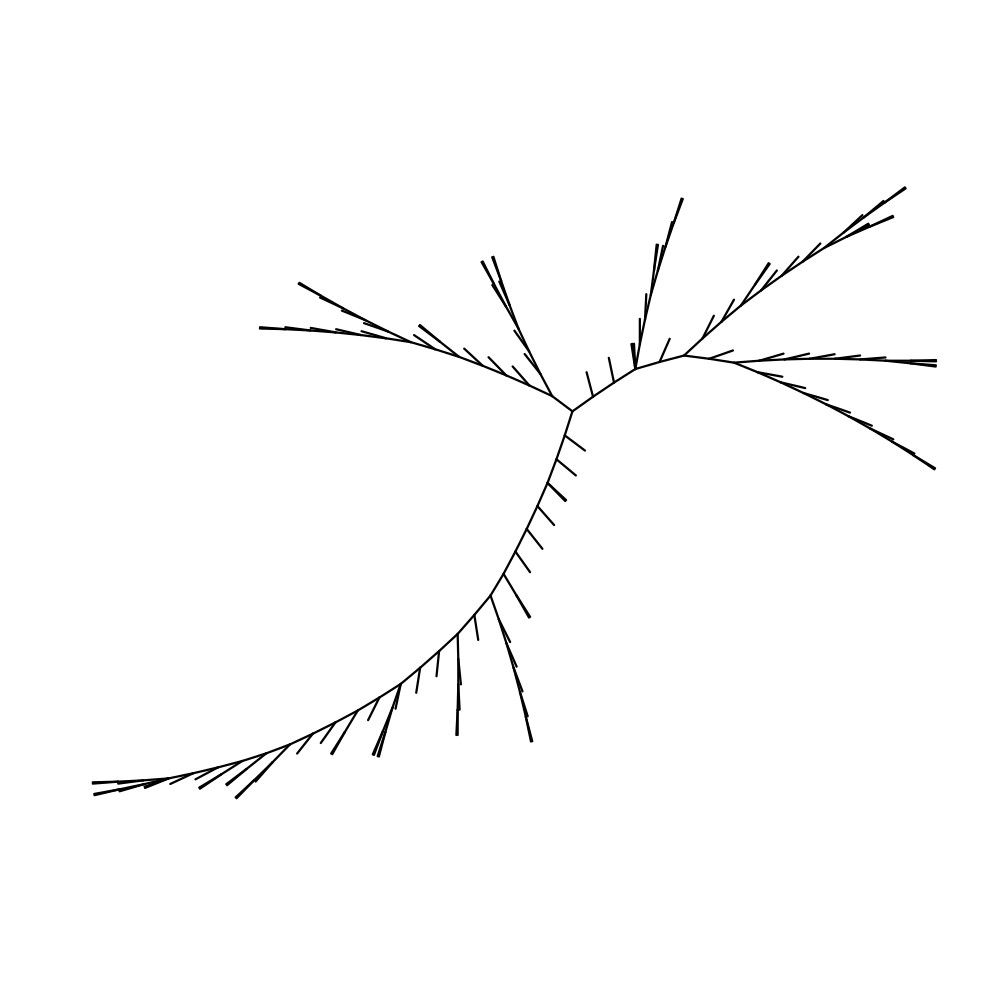}%
\end{center}
 \textbf{\refstepcounter{figure}\label{fig:clado} Figure \arabic{figure}.}{ A example tree obtained with cladistics, represented here as unrooted. When a root is chosen, the tree takes the shape of hierarchical trees. }
\end{figure}

Cladistics has been associated in the 80's to the search of a maximum parsimony tree. Maximum Parsimony is a powerful approach to find tree-like arrangements of objects (Fig.~\ref{fig:clado}). The drawback is that the analysis must consider all possible trees before selecting the most parsimonious one. The computation complexity depends on the number of objects and character states, so that too large samples (say more than a few thousands) cannot be analyzed.
The Maximum Parsimony algorithm can take uncertainties or unknowns into account by evaluating the different possibilities allowed by the range of values and selecting among them the one that provides the smallest score. In the case of unknown parameters, the most parsimonious diversification scenario provides a prediction for the unknown values.

In recent years the definition of cladistics has been extended to the classification of taxa (individuals or species) defined by characters on a rooted tree. In biological applications, a phylogenetic tree describes the possible evolution of a taxon corresponding to the root. The root may either be a real taxon or be inferred from the descendant taxa. The success of a cladistics analysis much depends on the behavior of the parameters. In particular, it is sensitive to redundancies, incompatibilities, too much variability (reversals), and parallel and convergent evolutions. It is thus a very good tool for investigating whether a given set of parameters can lead to a robust and pertinent diversification scenario.

If a set of characters exactly defines a phylogeny, then the phylogeny is called perfect.  In practical applications, the available characters seldom define a perfect phylogeny. A supplementary measure of the deviation to a perfect phylogeny is necessary to determine how well a candidate tree fits the characters. In the standard approach to parsimony, the score $s_p$ of a tree corresponds, after labeling of the internal nodes, to the minimum number of edges $(u,v)$ with $c(u)\neq c(v)$, $c(u)$ being the character state at node $u$. The tree with the minimum score is searched for with some heuristics \citep{Felsenstein1984}. The maximum parsimony approach can be directly extended to continuous characters or values. To each internal node is associated a real value $f(u)$. The score s of a tree equals the sum over all edges of the absolute difference between those values:

\begin{equation}
 \label{eq:MPs}
    s = \sum_{e=(u,v) \epsilon E}  \lvert f(u) - f(v) \rvert
 \end{equation}

\citet{robinson1973} has shown that for a tree defined by continuous characters, a maximum parsimony score is reached for values of the internal nodes belonging to the set of values (or states)  defined on the leaves.

The main method to search for the best tree representation of data beyond Maximum Parsimony include Maximum Likelihood. We note this technique which has never been applied to astrophysics in the context of classification but may be a pertinent approach. The problem here is that an evolutionary model must be used, and naturally the result will depend significantly on it. Maximum Likelihood is used standardly in biology, and it may be possible that astrophysicists could also have well constrained physical models of the evolution of galaxies and their properties.  The phylogenetic tree of Maximum Likelihood is the tree for which the observed data are most probable \citep{Williams2003}. Distance-based approaches are also often quite appropriate for reconstructing a phylogenetic tree from continuous characters. Distance-based approaches are fast and can be used for data exploration and for the selection of the most appropriate variables.

Cladistics when applied to domain outside of biology, like in astrocladistics, refers more generally to the classification of objects by a rooted or an unrooted tree (Fig.~\ref{fig:clado}). In that case, the tree represents possible relationships between taxa. The search of the best tree described by a set of characters on a set of objects (or taxa in phylogeny) can be done by several different approaches. The most popular methods are the one using Maximum Parsimony or Maximum Likelihood. For continuous parameters, the software program TNT \citep{TNT2008} is also quite popular to reconstruct trees from characters. As an alternative, the data can be discretized through appropriate binning.

As mentioned earlier, a new picture of phylogenies is emerging after the understanding that phylogenies on multistate characters reduce through a conceptually simple grouping of the characters into a phylogeny on binary characters. For binary characters, both distance- and character-based approaches are equivalent. This approach opens new perspectives as it furnishes also a bridge between character-based phylogenies and split networks or more precisely outer planar networks.

\textbf{Outer planar networks}

\begin{figure}[h!]
\begin{center}
\includegraphics[width=8cm]{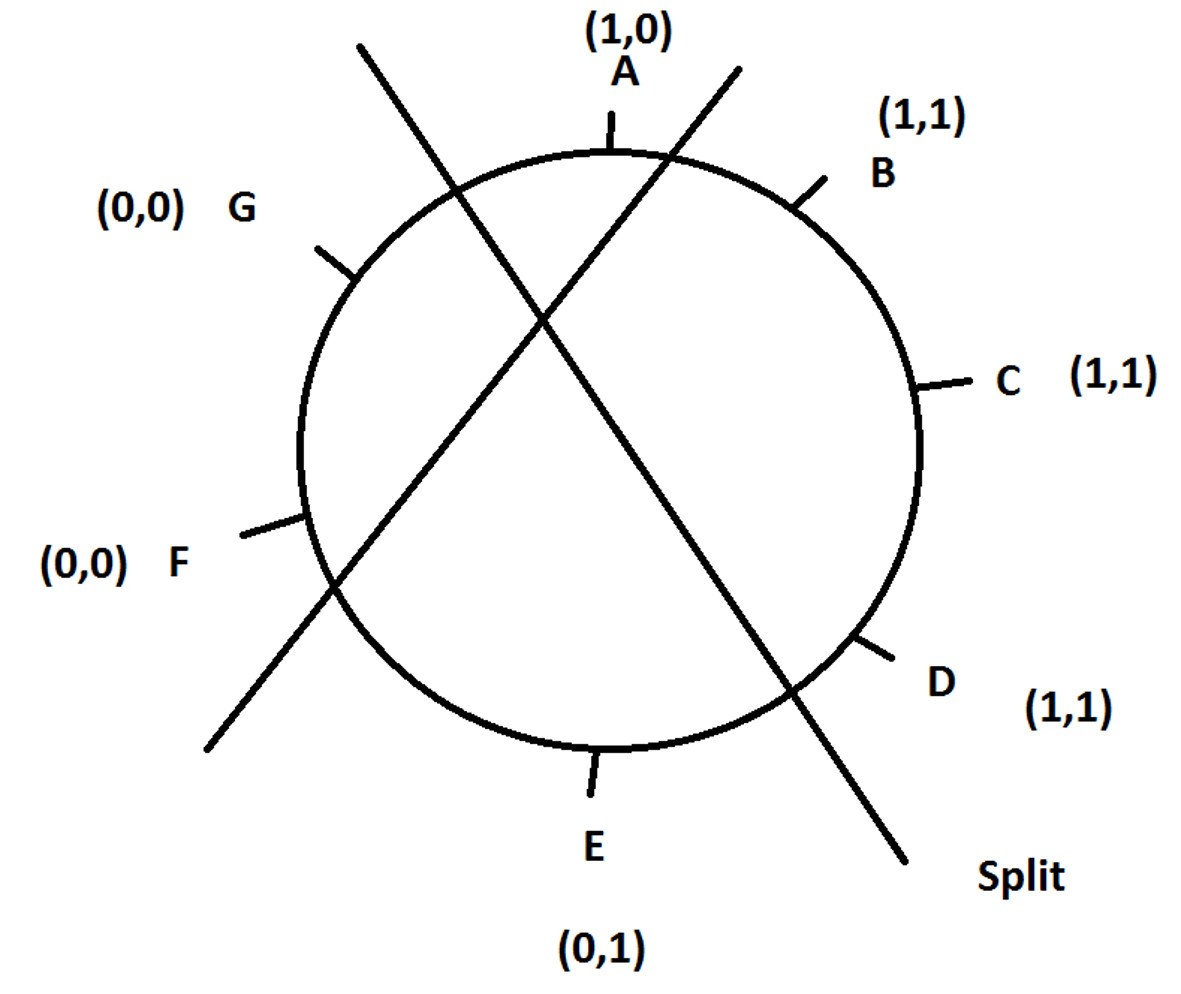}%
\end{center}
 \textbf{\refstepcounter{figure}\label{fig:split} Figure \arabic{figure}.}{ A circular order for objects A to G, with their pairs of binary states, arranged according to the circular consecutive-ones condition. The two lines show two different split, one between (0,*) and (1,*), the other one between (*,0) and (*,1).}
\end{figure}

Outer planar networks permit the simultaneous representation of alternative trees with reticulations, and are thus generalizations of trees \citep{Huson2006}. In order to understand the connection between outer planar networks and phylogenetic trees, one has to explain succinctly what is called a split on a circular order of the taxa.  A circular order on a phylogenetic tree corresponds to an indexing of the $n$ end nodes according to a circular (clockwise or anti-clockwise) scanning of the end nodes. A split on a circular order of the taxa is a partition of the objects into two disjoint sets (Fig.~\ref{fig:split}).

For multistate characters, a split can be defined after transformation of each multistate character into a binary character. For each pair of states (A,B), a subset of states containing the state ‘A’ is attributed the ‘1’ state and the complementary subset including the subset ‘B’ is given the binary state ‘0’. If the transformation can be done on each states and characters \citep[for details see][]{TF15} so that each binary character fulfills the circular consecutive-ones condition, then the data can be described exactly by an outer planar network. By definition the circular consecutive-ones condition are fulfilled if for any binary state, the taxa with the ‘1’ state are consecutive on the circular order (Fig.~\ref{fig:split}).

\begin{figure}[h!]
\begin{center}
\includegraphics[width=12cm]{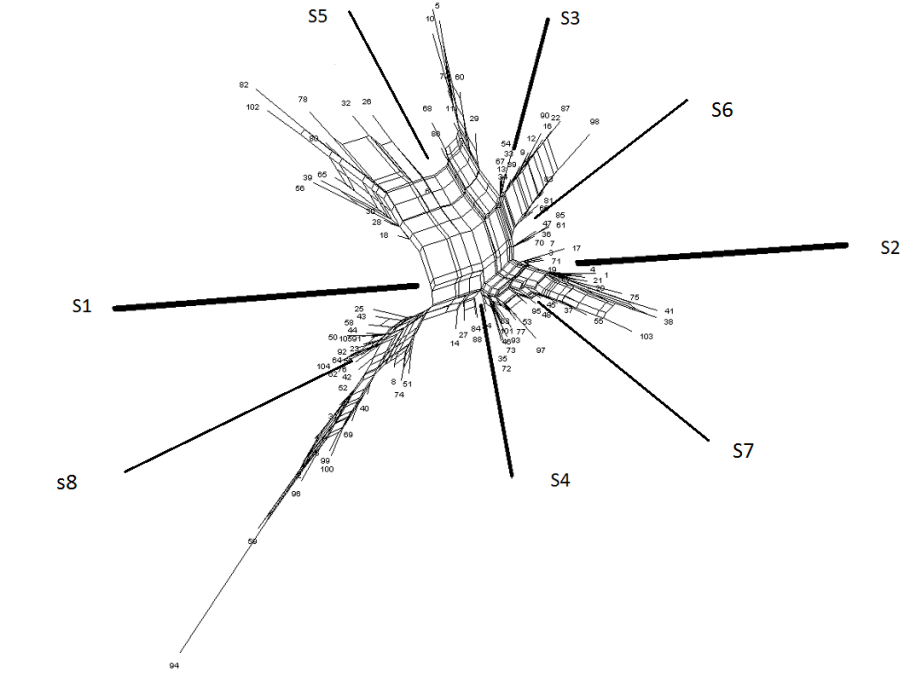}%
\end{center}
 \textbf{\refstepcounter{figure}\label{fig:splitnetwork} Figure \arabic{figure}.}{ An example of an outer planar network showing the eight splits of the eight parameters s1 ... s8.}
\end{figure}

Splits in an outer planar network (Fig.~\ref{fig:splitnetwork}) furnish neighboring relationships between objects. Objects sharing a common property, as defined by splits, are consecutive in a circular order. Outer planar networks can be regarded as a generalization of phylogenetic trees. An outer planar network reduces to a phylogenetic tree if for each pair of binary characters, the so-called 4-gamete rule is fulfilled. The 4-gamete rule states that for each pair of binary characters there is at least one of the 4 possible gametes (either (1.0), (0,1), (1,1) or (0,0)) that is missing.

For distance-based approach, the circular consecutive-ones conditions have to be replaced by the fulfillment of the Kalmanson inequalities.  For taxa indexed according to a circular order, the distances between
a reference node $n$ and the path $i-j$ are gathered in the distance matrix $\left\{Y_{i,j}^{n}\right\}$ with $Y_{i,j}^{n}=\frac{1}{2}\left(d_{i,n}+d_{j,n}-d_{i,j}\right)$, $d_{i,n}$ being the pairwise distance between leave $i$ and node $n$. This distance matrix fulfils the so-called Kalmanson inequalities:

\begin{equation}
 \label{eq:Kalmanson}
Y_{i,j}^{n} \geq Y_{i,k}^{n} \; , \; Y_{k,j}^{n}\geq Y_{k,i}^{n} \;\; (i\leq j\leq k)
\end{equation}

\citet{Bandelt1992} have shown  that  if  a  distance  matrix $\left\{Y_{i,j}^{n}\right\}$ fulfils Kalmanson inequalities, then the distance matrix can be exactly represented by a split network or by an X-tree.
The program SplitTrees4 \citep{Huson2006} permits to construct outer planar networks from a distance matrix.

In practice, the perfect order is not known or not feasible. The difference between the perfect order and the order one obtains with a given data set is called the contradiction. The minimum contradiction analysis \citep{Thuillard2007,Thuillard2008} finds the best order one can get. It is a powerful tool for ascertaining whether the parameters can lead to a tree-like arrangement of the objects \citep{TF09}. Using the parameters that fulfil this property, the method then performs an optimisation of the order and provides groupings with an assessment of their robustness.

We believe that outer planar networks will gain importance in applications outside of biology as they furnish a real alternative to the standard classification methods.


\subsection{Applications}

\citet{Farrah2009} have used a bayesian framework to compare and group 102 ultra-luminous infrared galaxy spectra and yield a network diagram which is used to define three groups. An evolutionary description of these galaxies is proposed from the properties of these groups. Even though their method is not a phylogenetic technique per se since the relationships are constructed after the clustering analysis, this work illustrates the potential need of phylogenetic tools in astrophysics.

The use of phylogenetic approaches in astrophysics has been pioneered and pursued through the denomination of astrocladistics \citep{jc1,jc2,FCD06}. Applications have been successfully performed for galaxies \citep{Fraix2010,Fraix2012}, globular clusters \citep{FDC09,FD15} and Gamma-Ray bursts \citep{Cardone2013}.

The phylogenetic approaches used on galaxy samples are clearly oriented towards a multivariate and evolutionary classification of galaxies \citep{Fraix2010,Fraix2012}. To this end, several statistical analyses (PCA, k-means, cladistics and minimum contradiction analysis) are used to select the set of parameters that yields a robust classification according to several clustering analyses (k-means, cladistics and Minimum Contradiction Analysis). Six parameters were so selected among the 23 available: the central velocity dispersion, the disc-to-bulge ratio, the effective surface brightness, the metallicity, and the line indices NaD and OIII. The agreement of the clustering obtained by different techniques reinforces largely the result. The cladistics tree (cladogram) is used for the interpretation since it also provides the relationships between the groups.

These relationships are hypothesized as being evolutionary so that the placement of the groups on some diagrams reflects the evolution of the properties and their correlations. For instance, the famous fundamental plane is not universal at all, this 3D correlation clearly depends on the diversification level of the group: the correlation becomes tighter when the history of a galaxy is more complex. Other well-known correlations, like Mg$_b$ vs the velocity dispersion, indeed disappear within the groups but is created by the alignment of the groups in the scatter plot. This strongly suggests that these correlations (known as scaling relations) are statistical and caused by a hidden confounding factor, which is possibly the evolution \citep{DFB2011}.

The new classification is rather easily interpreted with all the parameters available and by comparison with numerical simulations. The galaxies within a given group share a common history, that is a sequence of transforming events (collapse, interaction, harassment, merger...) that \citet{Fraix2012} are able to identify.

Outer planar or split networks have also been applied on galaxy samples \citep{TF09} even though it is for a theoretical illustration of an optimisation approach to fulfil as much as possible the Kalmanson inequalities (Eq.~\ref{eq:Kalmanson}). Nevertheless, a classification is obtained on this limited sample of 100 galaxies and with only three parameters. The main splitting character is the surface brightness (Brie) that separates the sample in two roughly equal bins.  Each branch is then split into two other branches defined by the character states, “low OIII”, “high OIII” for the “low Brie” branch and “low B-R”, “high B-R”  for the “high Brie” branch. The essential point here is that the split value separating ``low'' and ``high'' are not arbitrary at all, they are optimized according to an optimisation criterion aimed at obtaining the best split network or X-tree as possible.
Even though the result cannot be given too much generality due to the small sample, the astrophysical outcome is informative. First, all high  Brie  galaxies have high OIII, but some high OIII galaxies have low Brie. This means that some low  surface brightness galaxies in this sample have star formation, and some high surface brightness objects show  only  an  OIII  absorption  feature
Second, all high B-R galaxies have high Brie and high OIII. This means  that  in  this  sample, the red objects  have  a  high  surface  brightness and some star formation. They are thus not simply ageing galaxies, but probably form stars with high metallicity. Conversely, all low OIII galaxies of the sample have a low B-R, so that blue objects do not necessarily form a lot of stars.

\section{Conclusions and perspectives}

In the astrophysical literature, we have found that there is a growing interest for automated classification of galaxies, which is motivated mainly by the exploding amount of available data. For this purpose, more or less sophisticated statistical analyses are recognized to be necessary. In this paper, we have reviewed the techniques used so far. We do not claim to be exhaustive, but we think we have described quite a broad range of statistical tools.

Supervised learning analyses are mainly used to separate classes, morphological types or physical components in spectra of galaxies. The Principal Component Analysis is the most frequently used, due to its simplicity and efficiency, even though it is not a classification technique but rather a dimension reduction tool. Its attractiveness lies in its ability to perform automatic classification on moderately large data sets, and maybe more importantly, its ability to extract simple and important information from multivariate data. In this respect it greatly succeeds in separating spectra of galaxies, quasars and stars in large surveys.

The supervised learning approaches require a classification to be established beforehand. In nearly all cases, the traditional morphological classification is the reference. It thus appears that the astronomers are keen to devise an objective way of classifying galaxies, using modern tools and multivariate data, but the classes to retrieve are devised subjectively with a visual inspection of images in the visible, hence a rather monovariate source.

In the unsupervised learning analyses of the literature, the morphological classification also often serves as a reference that must be matched. However, many studies find different classes which bring new insights to the physics of galaxies and their evolution.  These classes are homogeneous in the multidimensional parameter space, and not necessarily in the traditional classification scheme. Because of the number of properties to consider, the description of these new classes is more complicated, but simpler (and more pertinent) when a comparison with models and numerical simulations is performed. In addition, new kinds of objects are found which would not be possible in a multidimensional parameter space with traditional approaches.

So an automatic classification of galaxies is becoming more and more crucial. The question remains of which classification is concerned. The predominance of the morphology as the most important parameter associated with the traditional classification scheme, is nearly overwhelming. Most unsupervised learning analyses yield new classifications, but this is not really exploited as such since their goal is often to propose an automatic way to retrieve the morphological classification.

We think that this goal is hopeless since it hides a fundamental contradiction between the classification obtained from a traditional visual subjective and monovariate approach and the one yielded by a multivariate automatic and objective technique. The fact that obvious correlations exist between the new classifications and the traditional one is a very strong support in favor of these advanced approaches and should not obliterate the difference in the classes.

The astrophysical results described in this review provide other arguments in favor of the statistical techniques, mainly because these tools can navigate more easily in a large dimensional space:

\begin{itemize}
 \item multivariate analyses are particularly interesting in the case of spectra, both for supervised and unsupervised classification. Dimension reduction is here an obvious requirement but proper unsupervised clustering is also necessary to discover new kinds of objects.
 \item for spectra, unsupervised techniques generally do not require fitting with model spectra, so that the comparison between models and observations can really be performed in the multivariate parameter space.
 \item more generally, the comparison between the observations, models and numerical simulations can be made by comparing the populations coming from the classifications of real and simulated galaxies, independently or together.
 \item soft (fuzzy), tree- or network-based classifications seem more appropriate to the continuous distribution of galaxy parameters than hard clustering.
 \item some techniques are based on the relationships between the objects and/or the classes. It is thus possible to objectively understand for instance the links between dynamically hot systems, or the place of the ``green valley'' galaxies with respect to ``blue'' and ``red'' ones, or the evolution of galaxies within the fundamental plane.
\end{itemize}

We conclude from this review that unsupervised analyses should not be afraid to propose new classifications of galaxies. These new classifications should be compared to other such classifications, this is the only way to draw a global view of galaxy diversity and be able to classify automatically galaxies of the present and future big surveys. In addition, and probably more importantly, the physics of galaxies being intrinsically multivariate, their classification cannot be based on only one criterion.

It is important to remember that there is not a unique best classification, and not a best tool. Comparison of results is a valuable task since it brings a lot of information on the nature of the data, the objects and their parameters. Also a classification is never definitive, and necessarily evolves with our knowledge and the discovery of new objects.

We wish to end this review with the cluster validation question. This is an important issue in clustering and classification. In general, cross-validation and bootstrap techniques are rather easy and provide good estimates of cluster robustness. Some other validation indices are Dunn’s Validation Index \citep{Dunn1973}, Davies-Bouldin Validity Index \citep{Davies1979}, C Index \citep{Hubert1976} and Silhoutte Validation Index \citep{Rousseeuw1987}. Many more are given in the $clusterCrit$ package of R.

In most of the clustering algorithms, the number of clusters are user specified. This is a difficult question, there are many tools (Sect.~\ref{kmeans}) to objectively guess the optimum number but they all have their drawbacks and limitations. Nevertheless, they should be used as much as possible to provide some hints and justifications.

\section*{Disclosure/Conflict-of-Interest Statement}

The authors declare that the research was conducted in the absence of any commercial or financial relationships that could be construed as a potential conflict of interest.

\section*{Author Contributions}

The contributions is mainly as follows: DFB performed the review of the astrophysical literature, MT developed the theoretical aspects of phylogenetic and wavelets methods, AKC developed the other statistical tool descriptions. All three authors equally participated to the elaboration of the documents.



%

\bibliographystyle{NewFrontiers}  
\bibliography{RevClustClass}

\begin{thebibliography}{90}
\expandafter\ifx\csname natexlab\endcsname\relax\def\natexlab#1{#1}\fi
\expandafter\ifx\csname urlstyle\endcsname\relax
  \expandafter\ifx\csname doi\endcsname\relax
  \def\doi#1{doi:\discretionary{}{}{}#1}\fi \else
  \expandafter\ifx\csname doi\endcsname\relax
  \def\doi{doi:\discretionary{}{}{}\begingroup \urlstyle{rm}\Url}\fi \fi
\expandafter\ifx\csname selectlanguage\endcsname\relax
  \def\selectlanguage#1{}\fi

\bibitem[{Albazzaz and Wang(2004)}]{Albazzaz2004}
Albazzaz H, Wang XZ.
\newblock Statistical process control charts for batch operations based on
  independent component analysis.
\newblock {\em Industrial \& engineering chemistry research\/} {\bf 43} (2004)
  6731--6741.

\bibitem[{{Allen} et~al.(2013){Allen}, {Hewett}, {Richardson}, {Ferland}, and
  {Baldwin}}]{Allen2013}
{Allen} JT, {Hewett} PC, {Richardson} CT, {Ferland} GJ, {Baldwin} JA.
\newblock Classification and analysis of emission-line galaxies using mean
  field independent component analysis.
\newblock {\em \mnras\/} {\bf 430} (2013) 3510--3536.
\newblock \doi{10.1093/mnras/stt151}.

\bibitem[{{Andrae} et~al.(2010){Andrae}, {Melchior}, and
  {Bartelmann}}]{Andrae2010}
{Andrae} R, {Melchior} P, {Bartelmann} M.
\newblock Soft clustering analysis of galaxy morphologies: a worked example
  with sdss.
\newblock {\em \aap\/} {\bf 522} (2010) A21.
\newblock \doi{10.1051/0004-6361/201014169}.

\bibitem[{{Ascasibar} and {Sanchez-Almeida}(2011)}]{Ascasibar2011}
{Ascasibar} Y, {Sanchez-Almeida} J.
\newblock Do galaxies form a spectroscopic sequence?
\newblock {\em \mnras\/} {\bf 415} (2011) 2417--2425.
\newblock \doi{10.1111/j.1365-2966.2011.18869.x}.

\bibitem[{{Ball} and {Brunner}(2010)}]{Ball2010}
{Ball} NM, {Brunner} RJ.
\newblock Data mining and machine learning in astronomy.
\newblock {\em International Journal of Modern Physics D\/} {\bf 19} (2010)
  1049--1106.
\newblock \doi{10.1142/S0218271810017160}.

\bibitem[{{Ball} et~al.(2007){Ball}, {Brunner}, {Myers}, {Strand}, {Alberts},
  {Tcheng} et~al.}]{Ball2007}
{Ball} NM, {Brunner} RJ, {Myers} AD, {Strand} NE, {Alberts} SL, {Tcheng} D,
  et~al.
\newblock Robust machine learning applied to astronomical data sets. {II}.
  quantifying photometric redshifts for quasars using instance-based learning.
\newblock {\em \apj\/} {\bf 663} (2007) 774--780.
\newblock \doi{10.1086/518362}.

\bibitem[{Bandelt and Dress(1992)}]{Bandelt1992}
Bandelt HJ, Dress AW.
\newblock Split decomposition: A new and useful approach to phylogenetic
  analysis of distance data.
\newblock {\em Molecular Phylogenetics and Evolution\/} {\bf 1} (1992) 242 --
  252.
\newblock \doi{http://dx.doi.org/10.1016/1055-7903(92)90021-8}.

\bibitem[{{Barrow} et~al.(1985){Barrow}, {Bhavsar}, and {Sonoda}}]{Barrow1985}
{Barrow} JD, {Bhavsar} SP, {Sonoda} DH.
\newblock Minimal spanning trees, filaments and galaxy clustering.
\newblock {\em \mnras\/} {\bf 216} (1985) 17--35.

\bibitem[{Bhavsar and Splinter(1996)}]{Bhavsar1996}
Bhavsar SP, Splinter RJ.
\newblock The superiority of the minimal spanning tree in percolation analyses
  of cosmological data sets.
\newblock {\em Monthly Notices of the Royal Astronomical Society\/} {\bf 282}
  (1996) 1461--1466.
\newblock \doi{10.1093/mnras/282.4.1461}.

\bibitem[{Boruvka(1926)}]{Boruvka1926}
Boruvka O.
\newblock O jistem problemu minimalnim (about a certain minimal problem).
\newblock {\em Praca Moravske Prirodovedecke Spolecnosti\/} {\bf 3} (1926)
  37--58.

\bibitem[{Cardone and Fraix-Burnet(2013)}]{Cardone2013}
Cardone F Vincenzo, Fraix-Burnet D.
\newblock {Hints for families of grbs improving the
  hubble diagram}.
\newblock {\em Monthly Notices of the Royal Astronomical Society\/} {\bf 434}
  (2013) 1930--1938.
\newblock \doi{10.1093/mnras/stt1122}.
\newblock 10 pages, 6 figures, 4 tables, accepted for publication on MNRAS.

\bibitem[{Charrad et~al.(2014)Charrad, Ghazzali, Boiteau, and
  Niknafs}]{NbClust}
Charrad M, Ghazzali N, Boiteau V, Niknafs A.
\newblock {NbClust}: An {R} package for determining the relevant number of
  clusters in a data set.
\newblock {\em Journal of Statistical Software\/} {\bf 61} (2014) 1--36.

\bibitem[{Chattopadhyay et~al.(2009)Chattopadhyay, Chattopadhyay, Davoust,
  Mondal, and Sharina}]{Chattopadhyay2009}
Chattopadhyay A, Chattopadhyay T, Davoust E, Mondal S, Sharina M.
\newblock Study of ngc 5128 globular clusters under multivariate statistical
  paradigm.
\newblock {\em The Astrophysical Journal\/} {\bf 705} (2009) 1533.

\bibitem[{Chattopadhyay et~al.(2013{\natexlab{a}})Chattopadhyay, Chattopadhyay,
  De, and Mondal}]{Chattopadhyay2013a}
Chattopadhyay AK, Chattopadhyay T, De T, Mondal S.
\newblock {\em Astrostatistical Challenges for the New Astronomy\/}
  (Springer-Verlag New York), {\em Springer Series in Astrostatistics\/},
  vol.~1, chap. Independent Component Analysis for Dimension Reduction
  Classification: Hough Transform and CASH Algorithm (2013{\natexlab{a}}),
  185--202.
\newblock \doi{10.1007/978-1-4614-3508-2}.

\bibitem[{Chattopadhyay et~al.(2013{\natexlab{b}})Chattopadhyay, Mondal, and
  Chattopadhyay}]{Chattopadhyay2013b}
Chattopadhyay AK, Mondal S, Chattopadhyay T.
\newblock Independent component analysis for the objective classification of
  globular clusters of the galaxy \{NGC\} 5128.
\newblock {\em Computational Statistics \& Data Analysis\/} {\bf 57}
  (2013{\natexlab{b}}) 17 -- 32.
\newblock \doi{http://dx.doi.org/10.1016/j.csda.2012.06.008}.

\bibitem[{Chattopadhyay and Chattopadhyay(2006)}]{Chattopadhyay2006}
Chattopadhyay T, Chattopadhyay A.
\newblock Objective classification of spiral galaxies having extended rotation
  curves beyond the optical radius.
\newblock {\em The Astronomical Journal\/} {\bf 131} (2006) 2452–2468.

\bibitem[{{Chattopadhyay} and {Karmakar}(2013)}]{Chattopadhyay2013}
{Chattopadhyay} T, {Karmakar} P.
\newblock Multivariate study of dynamically hot stellar systems: Clues to the
  origin of ultra compact and ultra faint dwarfs.
\newblock {\em New Astronomy\/} {\bf 22} (2013) 22--27.
\newblock \doi{10.1016/j.newast.2012.12.002}.

\bibitem[{Chattopadhyay et~al.(2007)Chattopadhyay, Misra, Naskar, and
  Chattopadhyay}]{ChattopadhyayGRB2007}
Chattopadhyay T, Misra R, Naskar M, Chattopadhyay A.
\newblock Statistical evidences of three classes of gamma ray bursts.
\newblock {\em The Astrophysical Journal\/} {\bf 667} (2007) 1017.

\bibitem[{Collet and Murtagh(2004)}]{Collet2004}
Collet C, Murtagh F.
\newblock Multiband segmentation based on a hierarchical markov model.
\newblock {\em Pattern Recognition\/} {\bf 37} (2004) 2337 -- 2347.
\newblock \doi{http://dx.doi.org/10.1016/j.patcog.2004.03.017}.

\bibitem[{Comon(1994)}]{Comon1994}
Comon P.
\newblock Independent component analysis, a new concept?
\newblock {\em Signal Processing\/} {\bf 36} (1994) 287 -- 314.
\newblock \doi{http://dx.doi.org/10.1016/0165-1684(94)90029-9}.
\newblock Higher Order Statistics.

\bibitem[{{Connolly} et~al.(1995){Connolly}, {Szalay}, {Bershady}, {Kinney},
  and {Calzetti}}]{Connolly1995}
{Connolly} AJ, {Szalay} AS, {Bershady} MA, {Kinney} AL, {Calzetti} D.
\newblock Spectral classification of galaxies: an orthogonal approach.
\newblock {\em \aj\/} {\bf 110} (1995) 1071.
\newblock \doi{10.1086/117587}.

\bibitem[{{Conselice}(2006)}]{Conselice2006}
{Conselice} CJ.
\newblock The fundamental properties of galaxies and a new galaxy
  classification system.
\newblock {\em \mnras\/} {\bf 373} (2006) 1389--1408.
\newblock \doi{10.1111/j.1365-2966.2006.11114.x}.

\bibitem[{{Coppa} et~al.(2011){Coppa}, {Mignoli}, {Zamorani}, {Bardelli},
  {Bolzonella}, {Pozzetti} et~al.}]{Coppa2011}
{Coppa} G, {Mignoli} M, {Zamorani} G, {Bardelli} S, {Bolzonella} M, {Pozzetti}
  L, et~al.
\newblock The bimodality of the 10k zcosmos-bright galaxies up to z \~{} 1: a
  new statistical and portable classification based on the global optical
  galaxy properties.
\newblock {\em \aap\/} {\bf 535} (2011) A10.

\bibitem[{{D'Abrusco} et~al.(2012){D'Abrusco}, {Fabbiano}, {Djorgovski},
  {Donalek}, {Laurino}, and {Longo}}]{DAbrusco2012}
{D'Abrusco} R, {Fabbiano} G, {Djorgovski} G, {Donalek} C, {Laurino} O, {Longo}
  G.
\newblock Clasps: A new methodology for knowledge extraction from complex
  astronomical data sets.
\newblock {\em \apj\/} {\bf 755} (2012) 92.
\newblock \doi{10.1088/0004-637X/755/2/92}.

\bibitem[{Davies and Bouldin(1979)}]{Davies1979}
Davies DL, Bouldin DW.
\newblock A cluster separation measure.
\newblock {\em Pattern Analysis and Machine Intelligence, IEEE Transactions
  on\/} {\bf PAMI-1} (1979) 224--227.
\newblock \doi{10.1109/TPAMI.1979.4766909}.

\bibitem[{{Davoodi} et~al.(2006){Davoodi}, {Oliver}, {Polletta},
  {Rowan-Robinson}, {Savage}, {Waddington} et~al.}]{Davoodi2006}
{Davoodi} P, {Oliver} S, {Polletta} MdC, {Rowan-Robinson} M, {Savage} RS,
  {Waddington} I, et~al.
\newblock Parametric modeling of the 3.6-8 {$\mu$}m color distributions of
  galaxies in the swire survey.
\newblock {\em \aj\/} {\bf 132} (2006) 1818--1833.
\newblock \doi{10.1086/506385}.

\bibitem[{De et~al.(2013)De, Chattopadhyay, and Chattopadhyay}]{De2013}
De T, Chattopadhyay T, Chattopadhyay AK.
\newblock Comparison among clustering and classification techniques on the
  basis of galaxy data.
\newblock {\em Calcutta Statistical Association Bulletin\/} {\bf 65} (2013)
  257--260.
\newblock \doi{10.1080/03610926.2013.848286}.
\newblock Special 8-th Triennial Proceedings Volume),.

\bibitem[{De et~al.(2014)De, Fraix-Burnet, and Chattopadhyay}]{De2014}
De T, Fraix-Burnet D, Chattopadhyay AK.
\newblock Clustering large number of extragalactic spectra of galaxies and
  quasars through canopies.
\newblock {\em Communication in Statistics - Theory and Methods\/}  (2014) in
  press.

\bibitem[{Dry et~al.(2009)Dry, Navarro, Preiss, and Lee}]{Dry09}
Dry M, Navarro D, Preiss K, Lee M.
\newblock The perceptual organization of point constellations.
\newblock {\em Annual Meeting of the Cognitive Science Society\/} (Amsterdam
  (The Netherlands)) (2009).

\bibitem[{Dunn(1973)}]{Dunn1973}
Dunn JC.
\newblock A fuzzy relative of the isodata process and its use in detecting
  compact well-separated clusters.
\newblock {\em Journal of Cybernetics\/} {\bf 3} (1973) 32--57.
\newblock \doi{10.1080/01969727308546046}.

\bibitem[{Fakcharoenphol et~al.(2003)Fakcharoenphol, Rao, and
  Talwar}]{Fakcharoenphol2003}
Fakcharoenphol J, Rao S, Talwar K.
\newblock A tight bound on approximating arbitrary metrics by tree metrics.
\newblock {\em Proceedings of the 35th Annual ACM Symposium on Theory of
  Computing\/} (San Diego, CA, USA) (2003), 448--455.
\newblock \doi{10.1.1.11.2667}.

\bibitem[{{Farrah} et~al.(2009){Farrah}, {Connolly}, {Connolly}, {Spoon},
  {Oliver}, {Prosper} et~al.}]{Farrah2009}
{Farrah} D, {Connolly} B, {Connolly} N, {Spoon} HWW, {Oliver} S, {Prosper} HB,
  et~al.
\newblock An evolutionary paradigm for dusty active galaxies at low redshift.
\newblock {\em \apj\/} {\bf 700} (2009) 395--416.
\newblock \doi{10.1088/0004-637X/700/1/395}.

\bibitem[{Fayyad et~al.(1996)Fayyad, Piatetsky-Shapiro, and Smyth}]{Fayyad1996}
Fayyad U, Piatetsky-Shapiro G, Smyth P.
\newblock From {Data} {Mining} to {Knowledge} {Discovery} in {Databases}.
\newblock {\em AI Magazine\/} {\bf 17} (1996) 37.
\newblock \doi{10.1609/aimag.v17i3.1230}.

\bibitem[{Feigelson and Babu(2012)}]{FeigelsonBabu2012}
Feigelson E, Babu G.
\newblock {\em Modern Statistical Methods for Astronomy: With R Applications\/}
  (Cambridge University Press) (2012).

\bibitem[{Felsenstein(1984)}]{Felsenstein1984}
Felsenstein J.
\newblock {\em Cladistics: Perspectives on the reconstruction of evolutionary
  history\/} (Columbia University Press, New York), chap. The statistical
  approach to inferring evolutionary trees and what it tells us about parsimony
  and compatibility (1984), 169--191.

\bibitem[{Felsenstein(2003)}]{Felsenstein2003}
Felsenstein J.
\newblock {\em Inferring Phylogenies\/} (Sinauer Associates, Sunderland,
  Massachusetts) (2003).

\bibitem[{{Folkes} et~al.(1996){Folkes}, {Lahav}, and {Maddox}}]{Folkes1996}
{Folkes} SR, {Lahav} O, {Maddox} SJ.
\newblock An artificial neural network approach to the classification of galaxy
  spectra.
\newblock {\em \mnras\/} {\bf 283} (1996) 651--665.

\bibitem[{Fraix-Burnet(2011)}]{DFB2011}
Fraix-Burnet D.
\newblock The fundamental plane of early-type galaxies as a confounding
  correlation.
\newblock {\em Monthly Notices of the Royal Astronomical Society: Letters\/}
  {\bf 416} (2011) L36--L40.
\newblock \doi{10.1111/j.1745-3933.2011.01091.x}.

\bibitem[{{Fraix-Burnet} et~al.(2012){Fraix-Burnet}, {Chattopadhyay},
  {Chattopadhyay}, {Davoust}, and {Thuillard}}]{Fraix2012}
{Fraix-Burnet} D, {Chattopadhyay} T, {Chattopadhyay} AK, {Davoust} E,
  {Thuillard} M.
\newblock A six-parameter space to describe galaxy diversification.
\newblock {\em {A}stronomy and {A}strophysics\/} {\bf 545} (2012) A80.
\newblock \doi{10.1051/0004-6361/201218769}.

\bibitem[{{Fraix-Burnet} et~al.(2006{\natexlab{a}}){Fraix-Burnet}, {C}holer,
  and {D}ouzery}]{FCD06}
{Fraix-Burnet} D, {C}holer P, {D}ouzery E.
\newblock {T}owards a {P}hylogenetic {A}nalysis of {G}alaxy {E}volution : a
  {C}ase {S}tudy with the {D}warf {G}alaxies of the {L}ocal {G}roup.
\newblock {\em {A}stronomy and {A}strophysics\/} {\bf 455} (2006{\natexlab{a}})
  845--851.
\newblock \doi{10.1051/0004-6361:20065098}.

\bibitem[{{Fraix-Burnet} et~al.(2006{\natexlab{b}}){Fraix-Burnet}, {C}holer,
  {D}ouzery, and {V}erhamme}]{jc1}
{Fraix-Burnet} D, {C}holer P, {D}ouzery E, {V}erhamme A.
\newblock {A}strocladistics: a phylogenetic analysis of galaxy evolution {I}.
  {C}haracter evolutions and galaxy histories.
\newblock {\em {J}ournal of {C}lassification\/} {\bf 23} (2006{\natexlab{b}})
  31--56.
\newblock \doi{10.1007/s00357-006-0003-5}.

\bibitem[{Fraix-Burnet and Davoust(2015)}]{FD15}
Fraix-Burnet D, Davoust E.
\newblock Stellar populations in $\omega$ centauri: a multivariate analysis.
\newblock {\em Monthly Notices of the Royal Astronomical Society\/} {\bf 450}
  (2015) 3431--3441.
\newblock \doi{10.1093/mnras/stv791}.

\bibitem[{{Fraix-Burnet} et~al.(2009){Fraix-Burnet}, {D}avoust, and
  {C}harbonnel}]{FDC09}
{Fraix-Burnet} D, {D}avoust E, {C}harbonnel C.
\newblock {T}he environment of formation as a second parameter for globular
  cluster classification.
\newblock {\em \mnras\/} {\bf 398} (2009) 1706--1714.
\newblock \doi{10.1111/j.1365-2966.2009.15235.x}.

\bibitem[{{Fraix-Burnet} et~al.(2006{\natexlab{c}}){Fraix-Burnet}, {D}ouzery,
  {C}holer, and {V}erhamme}]{jc2}
{Fraix-Burnet} D, {D}ouzery E, {C}holer P, {V}erhamme A.
\newblock {A}strocladistics: a phylogenetic analysis of galaxy evolution {II}.
  {F}ormation and diversification of galaxies.
\newblock {\em {J}ournal of {C}lassification\/} {\bf 23} (2006{\natexlab{c}})
  57--78.
\newblock \doi{10.1007/s00357-006-0004-4}.

\bibitem[{{Fraix-Burnet} et~al.(2010){Fraix-Burnet}, {Dugu{\'e}},
  {Chattopadhyay}, {Chattopadhyay}, and {Davoust}}]{Fraix2010}
{Fraix-Burnet} D, {Dugu{\'e}} M, {Chattopadhyay} T, {Chattopadhyay} AK,
  {Davoust} E.
\newblock Structures in the fundamental plane of early-type galaxies.
\newblock {\em \mnras\/} {\bf 407} (2010) 2207--2222.
\newblock \doi{10.1111/j.1365-2966.2010.17097.x}.

\bibitem[{Gascuel and Steel(2006)}]{NJ2006}
Gascuel O, Steel M.
\newblock Neighbor-joining revealed.
\newblock {\em Molecular Biology and Evolution\/} {\bf 23} (2006) 1997--2000.
\newblock \doi{10.1093/molbev/msl072}.

\bibitem[{Ghosh and Liu(2010)}]{kmeans2010}
Ghosh J, Liu A.
\newblock {\em The Top Ten Algorithms in Data Mining\/} (Taylor \& Francis),
  chap. The k-means Algorithm.
\newblock Chapman \& Hall/CRC Data Mining and Knowledge Discovery Series
  (2010), 21--36.

\bibitem[{Goloboff et~al.(2008)Goloboff, Farris, and Nixon}]{TNT2008}
Goloboff PA, Farris JS, Nixon KC.
\newblock {T}{N}{T}, a free program for phylogenetic analysis.
\newblock {\em Cladistics\/} {\bf 24.5} (2008) 774--786.

\bibitem[{Gower and Ross(1969)}]{Gower1969}
Gower JC, Ross GJS.
\newblock {Minimum spanning trees and single linkage
  cluster analysis}.
\newblock {\em Journal of the Royal Statistical Society. Series C (Applied
  Statistics)\/} {\bf 18} (1969) pp. 54--64.

\bibitem[{{Gratton} et~al.(2011){Gratton}, {Johnson}, {Lucatello}, {D'Orazi
  D'Orazi}, and {Pilachowski}}]{Gratton2011}
{Gratton} RG, {Johnson} CI, {Lucatello} S, {D'Orazi D'Orazi} V, {Pilachowski}
  C.
\newblock Multiple populations in {$\omega$} centauri: a cluster analysis of
  spectroscopic data.
\newblock {\em \aap\/} {\bf 534} (2011) A72.
\newblock \doi{10.1051/0004-6361/201117093}.

\bibitem[{Hennig(1965)}]{hennig1965}
Hennig W.
\newblock Phylogenetic systematics.
\newblock {\em Annual Review of Entomology\/} {\bf 10} (1965) 97--116.

\bibitem[{Hubert and Schultz(1976)}]{Hubert1976}
Hubert L, Schultz J.
\newblock Quadratic assignment as a general data analysis strategy.
\newblock {\em British Journal of Mathematical and Statistical Psychology\/}
  {\bf 29} (1976) 190--241.
\newblock \doi{10.1111/j.2044-8317.1976.tb00714.x}.

\bibitem[{{Huertas-Company} et~al.(2008){Huertas-Company}, {Rouan}, {Tasca},
  {Soucail}, and {Le F{\`e}vre}}]{HuertasCompany2008}
{Huertas-Company} M, {Rouan} D, {Tasca} L, {Soucail} G, {Le F{\`e}vre} O.
\newblock A robust morphological classification of high-redshift galaxies using
  support vector machines on seeing limited images. i. method description.
\newblock {\em \aap\/} {\bf 478} (2008) 971--980.
\newblock \doi{10.1051/0004-6361:20078625}.

\bibitem[{{Hurley} et~al.(2014){Hurley}, {Oliver}, {Farrah}, {Lebouteiller},
  and {Spoon}}]{Hurley2014}
{Hurley} PD, {Oliver} S, {Farrah} D, {Lebouteiller} V, {Spoon} HWW.
\newblock Learning the fundamental mid-infrared spectral components of galaxies
  with non-negative matrix factorization.
\newblock {\em \mnras\/} {\bf 437} (2014) 241--261.
\newblock \doi{10.1093/mnras/stt1875}.

\bibitem[{Huson and Bryant(2006)}]{Huson2006}
Huson DH, Bryant D.
\newblock Application of phylogenetic networks in evolutionary studies.
\newblock {\em Molecular Biology and Evolution\/} {\bf 23} (2006) 254--267.
\newblock \doi{10.1093/molbev/msj030}.

\bibitem[{Kaufman and Rousseeuw(1987)}]{kmedoids1987}
Kaufman L, Rousseeuw P.
\newblock Clustering by means of medoids.
\newblock Dodge Y, editor, {\em Statistical Data Analysis Based on the
  {L1}-Norm and Related Methods\/} (Elsevier/North-Holland, Amsterdam) (1987),
  405--416.

\bibitem[{Liu et~al.(2005)Liu, Duan, Liu, and Wu}]{Liu2005}
Liu R, Duan Fq, Liu Sy, Wu Fc.
\newblock Spectral classification of galaxy based on wavelet feature.
\newblock {\em ACTA ELECTRONICA SINICA\/} {\bf 33} (2005) 2059.
\newblock \doi{10.3321/j.issn:0372-2112.2005.11.031}.

\bibitem[{Lu et~al.(2006)Lu, Zhou, Wang, Wang, Dong, Zhuang et~al.}]{Lu2006}
Lu H, Zhou H, Wang J, Wang T, Dong X, Zhuang Z, et~al.
\newblock Ensemble learning for independent component analysis of normal galaxy
  spectra.
\newblock {\em The Astronomical Journal\/} {\bf 131} (2006) 790.

\bibitem[{MacQueen(1967)}]{kmeans1967}
MacQueen JB.
\newblock Some methods for classification and analysis of multivariate
  observations.
\newblock {\em Proceedings of 5th Berkeley Symposium on Mathematical Statistics
  and Probability\/} (1967), 281--297.

\bibitem[{Makarenkov et~al.(2006)Makarenkov, Kevorkov, and
  Legendre}]{Makarenkov2006}
Makarenkov V, Kevorkov D, Legendre P.
\newblock Phylogenetic network construction approaches.
\newblock Dilip K~Arora RMB, Singh GB, editors, {\em Bioinformatics\/}
  (Elsevier), {\em Applied Mycology and Biotechnology\/}, vol.~6 (2006), 61 --
  97.
\newblock \doi{http://dx.doi.org/10.1016/S1874-5334(06)80006-7}.

\bibitem[{{More} et~al.(2011){More}, {Kravtsov}, {Dalal}, and
  {Gottl{\"o}ber}}]{More2011}
{More} S, {Kravtsov} AV, {Dalal} N, {Gottl{\"o}ber} S.
\newblock The overdensity and masses of the friends-of-friends halos and
  universality of halo mass function.
\newblock {\em \apjs\/} {\bf 195} (2011) 4.
\newblock \doi{10.1088/0067-0049/195/1/4}.

\bibitem[{{Norman} et~al.(2004){Norman}, {Ptak}, {Hornschemeier}, {Hasinger},
  {Bergeron}, {Comastri} et~al.}]{Norman2004}
{Norman} C, {Ptak} A, {Hornschemeier} A, {Hasinger} G, {Bergeron} J, {Comastri}
  A, et~al.
\newblock The x-ray-derived cosmological star formation history and the galaxy
  x-ray luminosity functions in the chandra deep fields north and south.
\newblock {\em \apj\/} {\bf 607} (2004) 721--738.
\newblock \doi{10.1086/383487}.

\bibitem[{{Peth} et~al.(2015){Peth}, {Lotz}, {Freeman}, {McPartland},
  {Mortazavi}, {Snyder} et~al.}]{Peth2015}
{Peth} MA, {Lotz} JM, {Freeman} PE, {McPartland} C, {Mortazavi} SA, {Snyder}
  GF, et~al.
\newblock Beyond spheroids and discs: Classifications of candels galaxy
  structure at 1.4 $<$ z $<$ 2 via principal component analysis.
\newblock {\em ArXiv e-prints\/}  (2015).

\bibitem[{Qiu and Tamhane(2007)}]{Qiu2007}
Qiu D, Tamhane AC.
\newblock A comparative study of the k-means algorithm and the normal mixture
  model for clustering: Univariate case.
\newblock {\em Journal of Statistical Planning and Inference\/} {\bf 137}
  (2007) 3722 -- 3740.
\newblock \doi{http://dx.doi.org/10.1016/j.jspi.2007.03.045}.
\newblock Special Issue: In Celebration of the Centennial of The Birth of
  Samarendra Nath Roy (1906-1964).

\bibitem[{Reynolds et~al.(2006)Reynolds, Richards, Iglesia, and
  Rayward-Smith}]{kmedoids}
Reynolds A, Richards G, Iglesia B, Rayward-Smith V.
\newblock {Clustering rules: A comparison of
  partitioning and hierarchical clustering algorithms}.
\newblock {\em Journal of Mathematical Modelling and Algorithms\/} {\bf 5}
  (2006) 475--504.
\newblock \doi{10.1007/s10852-005-9022-1}.

\bibitem[{Robinson(1973)}]{robinson1973}
Robinson D.
\newblock Extending a function on a graph.
\newblock {\em Discrete Mathematics\/} {\bf 6} (1973) 89 -- 99.
\newblock \doi{10.1016/0012-365X(73)90038-1}.

\bibitem[{Rousseeuw(1987)}]{Rousseeuw1987}
Rousseeuw PJ.
\newblock Silhouettes: A graphical aid to the interpretation and validation of
  cluster analysis.
\newblock {\em Journal of Computational and Applied Mathematics\/} {\bf 20}
  (1987) 53 -- 65.
\newblock \doi{http://dx.doi.org/10.1016/0377-0427(87)90125-7}.

\bibitem[{Saitou and Nei(1987)}]{NJ1987}
Saitou N, Nei M.
\newblock The neighbor-joining method: a new method for reconstructing
  phylogenetic trees.
\newblock {\em Molecular Biology and Evolution\/} {\bf 4} (1987) 406--425.

\bibitem[{{S{\'a}nchez Almeida} et~al.(2010){S{\'a}nchez Almeida}, {Aguerri},
  {Mu{\~n}oz-Tu{\~n}{\'o}n}, and {de Vicente}}]{SanchezAlmeida2010}
{S{\'a}nchez Almeida} J, {Aguerri} JAL, {Mu{\~n}oz-Tu{\~n}{\'o}n} C, {de
  Vicente} A.
\newblock Automatic unsupervised classification of all sloan digital sky survey
  data release 7 galaxy spectra.
\newblock {\em ApJ\/} {\bf 714} (2010) 487--504.
\newblock \doi{10.1088/0004-637X/714/1/487}.

\bibitem[{{S{\'a}nchez Almeida} et~al.(2012){S{\'a}nchez Almeida}, {Terlevich},
  {Terlevich}, {Cid Fernandes}, and {Morales-Luis}}]{SanchezAlmeida2012}
{S{\'a}nchez Almeida} J, {Terlevich} R, {Terlevich} E, {Cid Fernandes} R,
  {Morales-Luis} AB.
\newblock Qualitative interpretation of galaxy spectra.
\newblock {\em \apj\/} {\bf 756} (2012) 163.
\newblock \doi{10.1088/0004-637X/756/2/163}.

\bibitem[{Sandage(2005)}]{Sandage2005}
Sandage A.
\newblock The classification of galaxies: Early history and ongoing
  developments.
\newblock {\em Annual Review of Astronomy \& Astrophysics\/} {\bf 43} (2005)
  581 -- 624.

\bibitem[{{Scarlata} et~al.(2007){Scarlata}, {Carollo}, {Lilly}, {Sargent},
  {Feldmann}, {Kampczyk} et~al.}]{Scarlata2007}
{Scarlata} C, {Carollo} CM, {Lilly} S, {Sargent} MT, {Feldmann} R, {Kampczyk}
  P, et~al.
\newblock Cosmos morphological classification with the zurich estimator of
  structural types (zest) and the evolution since z = 1 of the luminosity
  function of early, disk, and irregular galaxies.
\newblock {\em \apjs\/} {\bf 172} (2007) 406--433.
\newblock \doi{10.1086/516582}.

\bibitem[{{Semple} and {Steel}(2003)}]{semple2003}
{Semple} C, {Steel} MA.
\newblock {\em Phylogenetics\/} (Oxford: Oxford University Press) (2003).

\bibitem[{{Simpson} et~al.(2012){Simpson}, {Cottrell}, and
  {Worley}}]{Simpson2012}
{Simpson} JD, {Cottrell} PL, {Worley} CC.
\newblock Spectral matching for abundances and clustering analysis of stars on
  the giant branches of {$\omega$} centauri.
\newblock {\em \mnras\/} {\bf 427} (2012) 1153--1167.
\newblock \doi{10.1111/j.1365-2966.2012.22012.x}.

\bibitem[{{Slonim} et~al.(2001){Slonim}, {Somerville}, {Tishby}, and
  {Lahav}}]{Slonim2001}
{Slonim} N, {Somerville} R, {Tishby} N, {Lahav} O.
\newblock Objective classification of galaxy spectra using the information
  bottleneck method.
\newblock {\em \mnras\/} {\bf 323} (2001) 270--284.
\newblock \doi{10.1046/j.1365-8711.2001.04125.x}.

\bibitem[{Starck et~al.(1997)Starck, Siebenmorgen, and Gredel}]{Starck1997}
Starck JL, Siebenmorgen R, Gredel R.
\newblock Spectral analysis using the wavelet transform.
\newblock {\em The Astrophysical Journal\/} {\bf 482} (1997) 1011.

\bibitem[{{Suchkov} et~al.(2005){Suchkov}, {Hanisch}, and
  {Margon}}]{Suchkov2005}
{Suchkov} AA, {Hanisch} RJ, {Margon} B.
\newblock A census of object types and redshift estimates in the sdss
  photometric catalog from a trained decision tree classifier.
\newblock {\em \aj\/} {\bf 130} (2005) 2439--2452.
\newblock \doi{10.1086/497363}.

\bibitem[{Sugar and James(2003)}]{Sugar2003}
Sugar CA, James GM.
\newblock Finding the number of clusters in a dataset.
\newblock {\em Journal of the American Statistical Association\/} {\bf 98}
  (2003) 750--763.
\newblock \doi{10.1198/016214503000000666}.

\bibitem[{{Taghizadeh-Popp} et~al.(2012){Taghizadeh-Popp}, {Heinis}, and
  {Szalay}}]{TaghizadehPopp2012}
{Taghizadeh-Popp} M, {Heinis} S, {Szalay} AS.
\newblock Single parameter galaxy classification: The principal curve through
  the multi-dimensional space of galaxy properties.
\newblock {\em \apj\/} {\bf 755} (2012) 143.
\newblock \doi{10.1088/0004-637X/755/2/143}.

\bibitem[{Tajunisha and Saravanan(2010)}]{Tajunisha2010}
Tajunisha, Saravanan.
\newblock Performance analysis of k-means with different initialization methods
  for high dimensional data.
\newblock {\em International Journal of Artificial Intelligence \& Applications
  (IJAIA)\/} {\bf 1} (2010) 44--52.

\bibitem[{Team(2014)}]{R}
Team RC.
\newblock {\em R: A Language and Environment for Statistical Computing\/}.
\newblock R Foundation for Statistical Computing, Vienna, Austria (2014).

\bibitem[{Thuillard(2001)}]{Thuillard2001}
Thuillard M.
\newblock {\em Wavelets in Soft Computing\/}, {\em World Scientific Series in
  Robotics and Intelligent Systems\/}, vol.~25 (World Scientific) (2001).

\bibitem[{Thuillard(2007)}]{Thuillard2007}
Thuillard M.
\newblock Minimizing contradictions on circular order of phylogenic trees.
\newblock {\em Evolutionary Bioinformatics\/} {\bf 3} (2007) 267--277.
\newblock \doi{10.4137/EBO.S0}.

\bibitem[{Thuillard(2008)}]{Thuillard2008}
Thuillard M.
\newblock Minimum contradiction matrices in whole genome phylogenies.
\newblock {\em Evolutionary Bioinformatics\/} {\bf 4} (2008) 237--247.
\newblock \doi{10.4137/EBO.S909}.

\bibitem[{{T}huillard and {Fraix-Burnet}(2009)}]{TF09}
{T}huillard M, {Fraix-Burnet} D.
\newblock {P}hylogenetic {A}pplications of the {M}inimum {C}ontradiction
  {A}pproach on {C}ontinuous {C}haracters.
\newblock {\em {E}volutionary {B}ioinformatics\/} {\bf 5} (2009) 33--46.
\newblock \doi{10.4137/EBO.S2505}.

\bibitem[{{T}huillard and {Fraix-Burnet}(2015)}]{TF15}
{T}huillard M, {Fraix-Burnet} D.
\newblock A common framework for distance- and character- based phylogenies
  with applications to continuous characters.
\newblock {\em Evolutionary {B}ioinformatics\/}  (2015).
\newblock In revision.

\bibitem[{{Tishby} et~al.(2000){Tishby}, {Pereira}, and {Bialek}}]{Tishby2000}
{Tishby} N, {Pereira} FC, {Bialek} W.
\newblock The information bottleneck method.
\newblock {\em ArXiv Physics e-prints\/}  (2000).
\newblock Proceedings of the 37th Annu. Allerton Conference on Communications,
  Control, and Computing. Monticello, IL. pp. 368–377.

\bibitem[{Watanabe et~al.(1985)Watanabe, Kodaira, and Okamura}]{Watanabe1985}
Watanabe M, Kodaira K, Okamura S.
\newblock Digital surface photometry of galaxies toward a quantitative
  classification. iv - principal component analysis of surface-photometric
  parameters.
\newblock {\em The Astrophysical Journal\/} {\bf 292} (1985) 72--78.

\bibitem[{Whitmore(1984)}]{Whitmore1984}
Whitmore BC.
\newblock An objective classification system for spiral galaxies. {I} the two
  dominant dimensions.
\newblock {\em TheAstrophysical Journal\/} {\bf 278} (1984) 61--80.

\bibitem[{Williams and Moret(2003)}]{Williams2003}
Williams TL, Moret BM.
\newblock An investigation of phylogenetic likelihood methods.
\newblock {\em Bioinformatics and Bioengineering, 2003. Proceedings. Third IEEE
  Symposium on\/} (Bethesda, MD, USA: IEEE) (2003), 79--86.
\newblock \doi{10.1109/BIBE.2003.1188932}.

\end{thebibliography}

\end{document}